\documentclass[11pt,a4paper]{article}
\pdfoutput=1
\usepackage[]{inputenc,graphicx,amsmath,textcomp,amssymb,appendix}
\usepackage{subfigure}  % use for side-by-side figures
\newcommand{\be}{\begin{equation}}
\newcommand{\ee}{\end{equation}}
\newcommand{\bea}{\begin{eqnarray}}
\newcommand{\eea}{\end{eqnarray}}

\catcode`@=12

\begin{document}
\begin{flushright}
%SINP-APC-13/02
\end{flushright}
\thispagestyle{empty}
\begin{center}
{\Large\bf 
Possible Explanation of Indirect Gamma Ray Signatures from Hidden Sector
Fermionic Dark Matter}\\
\vspace{1cm}
{{\bf Amit Dutta Banik} \footnote{email: amit.duttabanik@saha.ac.in}, 
{\bf Debasish Majumdar} \footnote{email: debasish.majumdar@saha.ac.in}},
{\bf Anirban Biswas} \footnote{email: anirban.biswas@saha.ac.in}\\
\vspace{0.25cm}
{\normalsize \it Astroparticle Physics and Cosmology Division,}\\
{\normalsize \it Saha Institute of Nuclear Physics,} \\
{\normalsize \it 1/AF Bidhannagar, Kolkata 700064, India}\\
\vspace{1cm}
%%%%%%%%%%%%%%%%%%%%%%%%%%%%%%%%%%%%%%%%%%%%%%%%%%
%{\bf ABSTRACT}
%%%%%%%%%%%%%%%%%%%%%%%%%%%%%%%%%%%%%%%%%%%%%%%%%%
\end{center}
\begin{abstract}
We propose the existence of a hidden or dark sector besides
the standard model (SM) of particle physics, whose members (both
fermionic and bosonic) obey a local SU(2)$_{\rm H}$ gauge symmetry while
behaving like a singlet under the SM gauge group. However, the fermiomic fields
of the dark sector also possess another global U(1)$_{\rm H}$ symmetry
which remains unbroken. The local SU(2)$_{\rm H}$ invariance of
the dark sector is broken spontaneously when a scalar field in
this sector acquires a vacuum expectation value (VEV) and
thereby generating masses to the dark gauge bosons and dark
fermions charged under the SU(2)$_{\rm H}$. The lightest
fermion in this dark SU(2)$_{\rm H}$ sector can be a
potential dark matter candidate. We first examine the viability of the model
and constrain the model parameter space by theoretical
constraints such as vacuum stability and by the experimental
constraints such as PLANCK limit on relic density,
LHC data, limits on spin independent scattering cross-section
from dark matter direct search experiments etc.
We then investigate the gamma rays from the pair
annihilation of the proposed dark matter candidate at the Galactic Centre
region. We also extend our calculations of gamma rays flux
for the case of dwarf galaxies and compare the signatures
of gamma rays obtained from these astrophysical sites.
\end{abstract}
\newpage
\section{Introduction}
Experimental observations led by WMAP \cite{Hinshaw:2012aka}
and PLANCK \cite{Ade:2013zuv} satellites reveal that only about
4\% of our Universe is made up of ordinary baryonic matter and about
26.5\% of it is constituted by the unknown nonluminous matter or dark matter.
Dark matter (DM), is supposed to have a very weak interaction with
the visible sector of the Universe and strong evidence of its presence
come from gravitational probes only. However, the particle nature of DM
and reason for its high longevity (stability) are still unexplained.
Direct searches of DM are also performed by various DM direct
search experiments namely CDMS \cite{Agnese:2013cvt}-\cite{Agnese:2015ywx},
CoGent \cite{Aalseth:2010vx}, Xenon100 \cite{Aprile:2012nq},
LUX \cite{Akerib:2013tjd} etc. Although no convincing detection of DM has yet
been reported by these direct search experiments, until present, LUX experiment
provides most stringent bounds on DM spin independent elastic scattering
cross-sections $\sigma_{\rm SI}$ of the dark matter-nucleon with respect to its
mass.

Dark matter particles can get trapped inside massive astrophysical bodies
(due to their enormous gravity) like Galactic Centre (GC), the solar core. This
may happen if the velocities of DM particles fall below their escape velocities 
inside those massive bodies. When accumulated in considerable amount,
these trapped DM particles may undergo pair annihilation and
produce fermion-antifermion pairs, $\gamma$-rays etc.  
These $\gamma$-rays or fermions (positrons, neutrinos, antiprotons etc),
if found to be emitted in excess amount from these sites
which cannot be explained by known astrophysical processes,
may possibly be due to dark matter annihilation inside these sites. The
detection of such excess $\gamma$-rays, positrons, neutrinos etc. thus provide
valuable indirect signatures of the particle nature of
dark matter. Besides the GC, the dwarf galaxies
may also be rich is dark matter. The dwarf galaxies are a class of
faint and small satellite galaxies of our Milky Way galaxy.
The huge amount of DM content within these dwarf spherical galaxies (dsphs) is
inferred from their mass to luminosity ratio ($\frac {M} {L}$). The $\frac {M}
{L}$ ratio for these galaxies are found to be much higher than what is expected
from the estimation of their visible mass. The dark matter rich dsphs can also
emit excess $\gamma$-rays due to the pair annihilation of dark matter. Nine such
dwarf galaxies have recently been discovered in addition to the previously
discovered 15 dwarf  satellite galaxies of Milky way. Among these satellite
galaxies, the $\gamma$-ray flux obtained from the Reticulum 2 (Ret2 in brief)
dwarf galaxy shows an excess of $\gamma$-rays in the $\gamma$-ray energy range
of 2 to 10 GeV \cite{Geringer-Sameth:2015lua} while the null results from other
satellite galaxies provide stringent upper limits
on the annihilation cross-sections of DM particles for its various possible
annihilation channels \cite{Ackermann:2015zua, Drlica-Wagner:2015xua}.  
 
%The satellite borne experiments such as Fermi-LAT \cite{Ackermann:2012qk},
%AMS-02 \cite{Aguilar:2013qda} on board the International space station (ISS) as
%also ground based experiments such as H.E.S.S. \cite{Abramowski:2013ax} etc.
%look for such indirect signatures of dark matter as gamma, positron, anti-proton
%excess etc.
In the last few years, the analyses of Fermi-LAT publicly
available data \cite{fdata} by several groups
\cite{Goodenough:2009gk}- \cite{Abazajian:2014fta} have
confirmed the existence of a low energy (few GeV range) $\gamma$-ray excess
which appears to be emerging from the regions close to the centre of our Milky
way galaxy. The analysis of Fermi-LAT data \cite{fdata} by Daylan et. al.
\cite{Daylan:2014rsa} shows that the $\gamma$-ray excess from the GC can be
well explained
by the annihilation dark matter scenario. They have also excluded all the
known astrophysical processes which can act as the possible origin of
this phenomena. In Ref. \cite{Daylan:2014rsa} it is shown
that the observed $\gamma$-ray spectrum from the GC can
be well fitted by an annihilating dark matter particle
having mass in the range $\sim 30-40$ GeV which annihilates significantly
into ${b}\bar{b}$ final state with an annihilation cross-section
$\langle \sigma {\rm v}_{b\bar{b}}\rangle\sim(1.4-2.0)
\times10^{-26}{\rm cm}^3/{\rm s}$ with local dark matter density $\rho_\odot =
0.3$ GeV$/{\rm cm}^3$. In this work the authors have taken an angular region
of $5^0$ around the centre of our galaxy as their region of interest (ROI) and
used Navarro, Frenk and White (NFW) halo profile with $\gamma=1.26$ for the
computation of $\gamma$-ray flux. However more recently, the authors Calore,
Cholis and Weniger (CCW) of Ref. \cite{Calore:2014xka} have claimed to perform a
detailed analysis of Fermi-LAT data along with all the possible systematic
uncertainties using 60 galactic
diffusion excess (GDE) models. The results obtained from the analysis of CCW
provides a best fit for DM annihilation into $b \bar b$ final state
having mass $49^{+6.4}_{-5.4}$  GeV with  
$ {\langle \sigma {\rm v} \rangle}_{b \bar b} =
1.76^{+0.28}_{-0.27} \times 10^{-26}~{\rm cm}^3{\rm s}^{-1}$.
Moreover, the analysis of CCW assumes a NFW profile
with $\gamma=1.2$ and a different region of interest (ROI)
with galactic latitude $|l|\leq20^0$ and longitude $|b| \leq 20^0$
masking out the inner region corresponding to $|b| \leq2^0$.   
Different particle physics models for dark
matter that are simple extensions of Standard Model (SM) are proposed to
account for this 1-3 GeV excess in $\gamma$-rays from GC \cite{Boucenna:2011hy}-
\cite{Balazs:2015boa}.
It is to be noted that apart from dark matter, non-DM sources such as
millisecond pulsars may provide a feasible explanation to the excess of
$\gamma$-ray observed at GC \cite{Bartels:2015aea}. Study of unresolved point 
sources near GC by Lee et. al \cite{Lee:2015fea} suggests that point
sources also contribute significantly to the gamma ray excess. However, in this
work, we will consider DM as the origin of the observed excess in GC gamma ray to
explore the phenomenology of dark matter.  
Also the study of $\gamma$-rays from previously
known 15 different dwarf galaxies
by Fermi-LAT \cite{Ackermann:2015zua} and eight newly discovered dwarf
galaxies by Fermi-LAT with DES collaboration \cite{Drlica-Wagner:2015xua} give
bound on DM mass and corresponding $ {\langle \sigma {\rm v} \rangle}$ for
different annihilation channels.
   
Since the Standard Model (SM) of particle physics can not possibly provide for
the DM candidate, an extension of the SM is called for. Different particle
physics models for DM such as singlet scalar, fermion, vector where simple
extension of the SM with a
scalar, fermion or vector have been studied extensively in literature
\cite{Silveira:1985rk}-\cite{Ko:2014loa}. Inert doublet model (IDM)
\cite{Ma:2006km}-\cite{Modak:2015uda} also provides a viable DM candidate where an
extra Higgs doublet is considered along with the SM sector. Various extensions
of IDM with an additional scalar are also presented in Refs.
\cite{Banik:2014cfa,Bonilla:2014xba}. Several other models involving two
Higgs doublet model (THDM) accompanied by a singlet DM (scalar, fermion or
vector by choice) are also pursued in Refs. \cite{Aoki:2009pf}-\cite{Drozd:2014yla}. All
these models in Refs. \cite{Silveira:1985rk}-\cite{Drozd:2014yla} are based on a common
approach where DM particle is stabilised by assuming a discrete symmetry ($Z_2$ or $Z_3$)
and thus direct interactions (vertex with odd number of DM particle such
as decay term) with the SM fermions and gauge bosons are prohibited.
Hence, DM can interact with the visible sector only through the exchange of
Higgs or scalar bosons appearing in THDM. Also there are models with multi
component DM that presume discrete symmetry ($Z_2\times Z'_2$)
\cite{Biswas:2013nn} in order to stabilise the DM candidates. Different dark
matter models with continuous symmetry such as U(1) or SU(2) gauge symmetry are
also explored in literature \cite{Biswas:2014hoa},
\cite{Petriello:2008pu}-\cite{Ghorbani:2015baa}. 

In this work, we consider a ``hidden sector" framework
of dark matter without pretending any such discrete symmetry associated with it.
We propose the existence of a hidden sector which has SU(2)$_{\rm H}$ gauge structure.
Dark fermions in this hidden sector are charged under this SU(2)$_{\rm H}$ gauge group
while all the SM particles behave like a singlet.
Hence, the SM sector is decoupled from the dark sector
and could interact only through the exchange of scalar bosons that exist
in both the sector. Gauge bosons charged under SU(2)$_{\rm H}$ are
heavy and decay into dark fermions. Thus, the lightest one among dark fermions
is stable and can be treated as a viable DM candidate. 
%We check for the
%viability of the model by constraining the model parameter space by vacuum
%stability, LHC phenomenology, DM relic density, direct detection cross-section
%and also probe whether DM in present model can account for the indirect search
%results from GC and extragalactic $\gamma$-ray signals. 
We show that the DM candidate in the present model that satisfy
the limits from vacuum stability, LHC constraints, relic density,
direct detection experiments can duely explain the Galactic centre $\gamma$-ray
excess and also is in agreement with the limits on DM annihilation cross-section
obtained from the study of dwarf galaxies. 

The paper is organised as follows : In Sect.~\ref{S:model} we present the 
hidden sector SU(2)$_{\rm H}$ model. Constraints and bound on the model 
parameter space from vacuum stability, LHC results on SM Higgs, relic density
etc. are described in Sect.~\ref{S:cons}. In Sect.~\ref{S:res} we show the
results obtained for the available model parameter space and study of indirect
searches of $\gamma$-ray is performed. Finally in Sect.~\ref{S:con} we summarise
the work with concluding remarks. 
\section{The Model}
\label{S:model}
We consider the existence of a ``dark sector" that governs the particle
candidate of dark matter. Just as the ``visible sector" related to the known
fundamental particles successfully explained by the Standard Model, we propose
the existence of a hidden ``dark sector" that relates the dark matter particles.
We also presume that the Lagrangian of this hidden sector remains invariant
under the transformations of a local SU(2)$_{\rm H}$ as well as a global U(1)$_{\rm H}$
gauge symmetries. Therefore we consider two fermion generations
$\chi_{_{_{1}}}$ ($i=1,\,2$) where each generation consists of two fermions.
Consequently, in the dark sector we have altogether four
fermions namely $f_i$ ($i=1,4$). The left handed component of each
fermion (${f_i}_{\rm L}$) transforms like a part of a doublet
under SU(2)$_{\rm H}$ while its right handed part ${f_i}_{\rm R}$
behaves like a singlet under the same gauge group. Thus, the left handed
components of $f_1$, $f_2$ and $f_3$, $f_4$ form two separate
SU(2)$_{\rm H}$ doublets\footnote{In order to cancel the Witten anomaly
\cite{Witten:1982fp} we need at least two (even numbers) of left handed
fermionic SU(2)$_{\rm H}$ doublets in our model.}. However, both
the left handed as well as the right handed fermionic components
are charged under the postulated global U(1)$_{\rm H}$ symmetry.
The interactions between the dark sector fermions and the
SM particles are possible by the presence of an SU(2)$_{\rm H}$
scalar doublet $\Phi$ through the gauge invariant interaction term
$\lambda_3 H^\dagger H \Phi^\dagger \Phi$ which introduces
a finite mixing between the SM Higgs boson and the neutral
component of the hidden sector scalar doublet $\Phi$.
This scalar doublet does not have any global U(1)$_{\rm H}$
charge. As a result, the global U(1)$_{\rm H}$ symmetry
does not break spontaneously. However, being an SU(2)$_{\rm H}$
doublet $\Phi$ breaks the local SU(2)$_{\rm H}$
symmetry spontaneously when its neutral component
acquires vacuum expectation value (VEV) $v_s$.
Besides the local SU(2)$_{\rm H}$
gauge symmetry, the scalar doublet $\Phi$,
which is in the fundamental representation of SU(2)$_{\rm H}$
gauge group, also possesses a custodial SO(3) symmetry.
As a result of this residual SO(3) symmetry, three dark gauge bosons
${A_{i}^{\prime}}_{\mu}$ ($i=1$ to 3) which get mass due to
the spontaneous breaking of the local SU(2)$_{\rm H}$ symmetry,
become degenerate in mass. Non abelian nature of the SU(2)$_{\rm H}$
forbids the mixing between SM gauge bosons with dark gauge bosons
${A_{i}^{\prime}}_{\mu}$ ($i=1$ to 3) \cite{Hambye:2008bq,DiChiara:2015bua}.
The scalar doublets $H$, $\Phi$ and the fermionic doublets can be written 
as\footnote{Although, in order to keep similarity with the expression of the
Standard Model Higgs doublet $H$, we have introduced the notation of three
scalar fields, in the expression of $\Phi$, as $G_2^+,~\phi^0$ and $G_2^0$,
however the symbols $+$ and $0$ appearing in the superscript of dark sector
scalar fields do not represent the electric charge of the corresponding scalar
field as electric charge itself is not defied in the dark sector which is
invariant only under SU(2)$_{\rm H}$.}
\begin{eqnarray}
H = \left(\begin{array}{cc} G_1^+ \\
\frac{h^0+iG_1^0}{\sqrt{2}}\end{array}\right)\,,\,\,\,
\Phi = \left(\begin{array}{cc}G_2^+ \\
\frac{\phi^0+iG_2^0}{\sqrt{2}}\end{array}\right)\,,\,\,\,
\chi_{_{_{1 \rm L}}} = \left(\begin{array}{cc} f_1 \\
f_2\end{array}\right)_{{}_{\rm L}}\,,\,\,\,
\chi_{_{_{2 \rm L}}} = \left(\begin{array}{cc} f_3 \\
f_4\end{array}\right)_{{}_{\rm L}}\,\,.
\label{prob1}
\end{eqnarray}
%which is manifested by the presence of three massive gauge fields .
Therefore, the most general Lagrangian of the present proposed model
contains the following gauge invariant terms 
\begin{eqnarray}
\mathcal{L} &\supset& -\frac{1}{4} F^{\prime}_{\mu \nu} F^{\prime \mu \nu}
+ (D_{\mu} H)^\dagger (D^{\mu} H) + (D^\prime_{\mu} \Phi)^\dagger (D^{\prime\mu}\Phi)
- \mu_1^2~H^\dagger H  - \mu_2^2~\Phi^\dagger \Phi
\nonumber \\&& 
-\lambda_1 (H^\dagger H)^2 - \lambda_2 (\Phi^\dagger \Phi)^2
- \lambda_3~H^\dagger H \Phi^\dagger \Phi 
+ \sum_{i=1,2}\bar{\chi}_{_{_{i}}}{_{_{\rm L}}}
(i{D^\prime\!\!\!\!\!\slash}\,\,\chi_{_{_{i}}}{_{_{\rm L}}})
%\nonumber \\&&
+\sum_{i=1,4} \bar{f}_{i\,\,\rm R}(i\partial\!\!\!\slash\,f_{i\,\,\rm R})
\nonumber \\&&
-y_{1}^\prime\,\bar{\chi}_{_{_{1}}}{_{_{\rm L}}} \Phi f_{1\,\,\rm R}
-y_{2}^\prime\,\bar{\chi}_{_{_{1}}}{_{_{\rm L}}} \tilde{\Phi} f_{2\,\,\rm R},
-y_{3}^\prime\,\bar{\chi}_{_{_{2}}}{_{_{\rm L}}} \Phi f_{3\,\,\rm R}
-y_{4}^\prime\,\bar{\chi}_{_{_{2}}}{_{_{\rm L}}} \tilde{\Phi} f_{4\,\,\rm R}\,+\,\,hc \,\,\, ,
\label{prob2}
\end{eqnarray}
with 
\begin{eqnarray}
D_{\mu} &=& (\partial_{\mu} + i \frac{g}{2}\sum_{a = 1, 3}
\sigma_{a}{W^{a}}_{\mu} +i \frac{g^\prime}{2}B_{\mu}) \nonumber \,\,, \\
D^\prime_{\mu} &=& (\partial_{\mu} + i \frac{g_{{}_{\rm H}}}{2}
\sum_{a = 1, 3}\sigma^{a}{A_{a}^\prime}_{\mu})\,\,,
\label{prob3}
\end{eqnarray}
are the covariant derivatives of the SU(2)$_{\rm L}\times{\rm U(1)_{Y}}$ doublet
$H$ and the SU(2)$_{\rm H}$ doublets $\Phi$, $\chi_{_{_{i \rm L}}}$
respectively while $\tilde{\Phi}=i\sigma_2 \Phi^\star$ with $\sigma_2$ is the
Pauli spin matrix. Moreover, $g$, $g^\prime$ and $g_{{}_{\rm H}}$ are the respective gauge couplings 
corresponding to the gauge groups SU(2)$_{\rm L}$, U(1)$_{\rm Y}$ and
SU(2)$_{\rm H}$. In the above
equation (Eq.~\ref{prob2}) $F^\prime_{\mu \nu}$ is the field strength
tensor for the gauge fields ${A_{i}^{\prime}}_{\mu}$ ($i=1$ to 3) of the
SU(2)$_{\rm H}$ gauge group while $H$ is the usual SM Higgs doublet.
The global U(1)$_{\rm H}$ invariance of the dark sector Lagrangian
forbids the presence of any Majorana type mass terms of the fermionic
fields ($f_i$, $i=1$, 4) in Eq. \ref{prob2}.
We have assumed at the beginning that the dark sector
fermions are charged under a global U(1)$_{\rm H}$ symmetry.
Therefore invariance of the dark sector Lagrangian (Eq. \ref{prob2})
under this U(1)$_{\rm H}$ symmetry requires an equal and opposite U(1)$_{\rm H}$
charges between each fermion and its antiparticle. Thus we can say that
there is some conserved quantum number in the theory which can differentiate
between a fermion and its antiparticle. In other words this can be stated
the dark sector fermions in the present theory are Dirac type fermions.
We have also assumed that the dark sector fermions ($f_i$, $i=1,\,4$)
are in ``mass basis'' or ``physical basis'' so that the
Lagrangian (Eq. \ref{prob2}) does not contain any mixing
term between these fermionic states\footnote{ Alternatively, one may think
that the fermions in dark sector may have mixing between themselves similar to
the case of SM fermions in quark and lepton sectors. Following the CKM mechanism
in the quark sector of SM we can assume that the mass matrix of up-type fermion
generations i.e. $f_1$ and $f_3$ is diagonal while the mixing takes place
between down-type fermionic states ($f_2$, $f_4$). Now, since we have considerd
the up-type fermion $f_1$ to be the lightest of all fermions in dark sector,
thus in the present framework, the study of fermion mixing is redundant.}.
The dark sector fermions can interact among themselves by exchanging
dark gauge bosons ${A_{i}^{\prime}}_{\mu}$ and due to the presence of
these interaction modes all the heavier fermions such as $f_i$ ($i=2$ to 4)
decay into the lightest one ($f_1$). Consequently, the lightest fermion
$f_{1}$ is stable and can be a viable dark matter candidate. Like the
hidden sector gauge fields $A^\prime_{i \mu}$, the dark matter candidate
$f_1$ also gets mass when the postulated SU(2)$_{\rm H}$ symmetry of the
hidden sector breaks spontaneously by the VEV of $\Phi$. Thus, the
expression of mass of the fermionic dark matter candidate can easily be
obtained using Eq.~\ref{prob2} which is    
\begin{equation}
m_{f_{1}} = \frac{y^\prime_1 v_s}{\sqrt{2}}.
\label{prob4}
\end{equation}

We have already mentioned before, that due to the presence of the gauge
invariant term $\lambda_3 H^\dagger H \Phi^\dagger \Phi$, the neutral
components of both the scalar doublets, namely $h^0$ and $\phi^0$,
possess mass mixing between themselves. The mass squared mixing matrix
between these two real scalar fields are given by,
\begin{eqnarray}
\mathcal{M}^2_{\rm scalar} = \left(\begin{array}{cc}
2\lambda_1 v^2 ~~&~~ \lambda_3 v v_s \\
~~&~~\\
\lambda_3 v v_s ~~&~~ 2 \lambda_2 v_s^2
\end{array}\right) \,\,.
\label{prob5}
\end{eqnarray}
After diagonalising the mass squared matrix $\mathcal{M}^2_{\rm scalar}$,
we obtain two physical eigenstates $h_1$ and $h_2$ which are related to the
old basis sates $h^0$ and $\phi^0$ by an orthogonal transformation matrix
$O(\alpha)$ where $\alpha$ is the mixing angle between the resulting physical
scalars. The relation between physical scalars $h_1$ and $h_2$ with the 
scalar fields $h^0$ and $\phi^0$ are given as
\begin{eqnarray}
h_1=\cos\alpha~h^0 - \sin\alpha~\phi^0 \, ,\nonumber \hskip 15 pt
h_2=\sin\alpha~h^0 + \cos\alpha~\phi^0 \, .
\label{prob6}   
\end{eqnarray} 
The expressions of the mixing angle $\alpha$ and the masses of
the physical real scalars $h_1$ and $h_2$ are given by 
\begin{eqnarray}
\alpha &=& \frac{1}{2}~\tan^{-1}\left(\frac{\frac{\lambda_3}{\lambda_2}\frac{v}{v_s}}
{1 - \frac{\lambda_1}{\lambda_2}\frac{v^2}{v^2_s}}\right) \,\, ,\\
m_1 &=& \sqrt{\lambda_1 v^2 + \lambda_2 v^2_s + 
\sqrt{(\lambda_1 v^2 - \lambda_2 v^2_s)^2 + (\lambda_3 v v_s)^2} }\ ,\nonumber \\
m_2 &=& \sqrt{\lambda_1 v^2 + \lambda_2 v^2_s - 
\sqrt{(\lambda_1 v^2 - \lambda_2 v^2_s)^2 + (\lambda_3 v v_s)^2} } \,\ .
\label{prob7}
\end{eqnarray}
We assume the physical scalar $h_1$ is the SM-like Higgs boson which
has been observed by the ATLAS and the CMS detector 
\cite{Aad:2012tfa,Chatrchyan:2012ufa}. Therefore we have adopted the mass
($m_1$) of $h_1$ and VEV $v$ of $h^0$ to be $\sim 125.5$ GeV and 246 GeV
respectively. Thus, we have three unknown model parameters which control the
interactions of the dark matter candidate $f_1$ in the early Universe, namely
the mixing angle $\alpha$, the mass ($m_2$) of the extra physical scalar boson
$h_2$ and more importantly, the mass $m_{f_1}$ of the dark matter particle
$f_1$. In the rest of our work we have computed the allowed ranges of these
model parameters using various theoretical, experimental as well as
observational results. Throughout the work, for simplicity we take mass of
fermionic DM candidate ($f_1$) to be $m$.      
\section{Constraints}
\label{S:cons}
In this section we will discuss various constraints and bounds on model
parameters that arise from both theoretical aspects and experimental 
observations.
\begin{itemize}
\item {\bf Vacuum Stability} - 
To ensure the stability of the vacuum, the scalar potential for the model must
remain bounded from below. The quartic terms of the scalar potential is given 
as
\bea
V_4 &=& \lambda_1 (H^\dagger H)^2 + \lambda_2 (\Phi^\dagger \Phi)^2 + \lambda_3
~H^\dagger H \Phi^\dagger \Phi \,\, ,
\label{1}
\eea
where $H$ is the SM Higgs doublet and $\rm \Phi$ is the hidden sector Higgs
doublet. Conditions for the vacuum stability in this framework is given as
\bea
\lambda_1 > 0,\hskip 15pt \lambda_2 > 0\, ,\hskip 15 pt   
\lambda_3 + 2\sqrt{\lambda_1\lambda_2}  >  0\,\, .
\label{2}
\eea
\item {\bf LHC Phenomenology} -
In the present model of hidden sector (SU(2)$_{\rm H}$) fermionic dark matter
discussed earlier in Sect.~\ref{S:model}, an extra Higgs doublet is added to the
SM. This dark SU(2)$_{\rm H}$ Higgs doublet provides an additional Higgs like
scalar
that mixes up with the SM Higgs. Large Hadron Collider (LHC) performing the
search of Higgs particle (ATLAS and CMS Collaboration) have already discovered a
Higgs like particle having mass about 125 GeV. The excess in $\gamma\gamma$ 
channel reported independently by ATLAS \cite{Aad:2012tfa} and CMS
\cite{Chatrchyan:2012ufa}
confirmed the existence of Higgs like bosons. In the case of Hidden sector
SU(2)$_{\rm H}$ model, the mixing
between SM Higgs with Dark Higgs results in two Higgs like scalars. In the
the present scenario we take one of the scalar $(h_1)$ as the SM Higgs with 
mass $m_1= 125$ GeV. We further assume that the signal strength of scalar $h_1$
also satisfies the limits on the same obtained for the newly discovered boson.
Thus, $h_1$ in the present framework is identical with the SM like Higgs as 
reported by LHC Higgs search experiments (ATLAS and CMS). The signal
strength of Higgs boson ($h$), decaying into a particular final state ($xx$, $x$
is any SM particle), is defined as
\bea
R &=& \frac {\sigma (pp\rightarrow h)} {\sigma^{\rm SM}(pp\rightarrow h)}
\frac {{\rm Br} (h\rightarrow xx)} {{\rm Br}^{\rm SM}(h\rightarrow xx)} \,\,,
\label{3}
\eea     
where $\sigma(pp\rightarrow h)$ and $\rm Br(h\rightarrow x x)$
are the Higgs production cross-section and its branching ratio of any particular
decay mode ($x=$quark, lepton or gauge boson),
obtained from LHC experiments. The corresponding quantities
computed using Standard Model of electroweak interaction are
denoted by $\sigma^{\rm SM}(pp\rightarrow h)$ and
${\rm Br}^{\rm SM}(h\rightarrow xx)$ respectively.
For the present model, the signal strength of the SM-like scalar $h_1$
is then defined as,
\bea
R_1 =  \frac {\sigma (pp\rightarrow h_1)} {\sigma^{\rm SM}(pp\rightarrow h)}
\frac {{\rm Br} (h_1 \rightarrow xx)} {{\rm Br}^{\rm SM}(h \rightarrow xx)} \,\, , 
\label{4}
\eea  
where the quantities are in the numerator of Eq. \ref{4}
are the production cross-section and branching ratio of SM-like
Higgs boson $h_1$ which are computed using the present formalism.
Now due to the mixing of scalar bosons, the coupling of
SM-like Higgs boson to the SM fermions and gauge bosons
are modified with respect to SM Higgs boson ($h$) by
the cosine of mixing angle $\alpha$ whereas, the couplings
of non-SM scalar boson $h_2$ to SM particles are multiplied
by a factor $\sin \alpha$. Hence
the ratio $\frac {\sigma (pp\rightarrow h_1)} {\sigma^{\rm SM}(pp\rightarrow
h)} = \cos^2 \alpha$ and from the similar argument one can yield
$\frac {\sigma (pp\rightarrow h_2)} {\sigma^{\rm SM}(pp\rightarrow
h)}=\sin^2 \alpha$. The SM branching ratio can be expressed as 
${\rm Br}^{\rm SM} (h\rightarrow xx)= \frac{\Gamma^{\rm SM} (h\rightarrow
xx)}{\Gamma^{\rm SM}}$ where $\Gamma^{\rm SM} (h\rightarrow xx)$ is the
decay width of SM Higgs boson $h$ into any final state particles
and ${\Gamma^{\rm SM}}$ is the total SM Higgs decay width
having mass $m_1=125$ GeV. Similarly one can derive the expression
for branching ratio of $h_1$ into any specific decay channel in
the present model ${\rm Br}(h_1\rightarrow xx)= \frac{\Gamma_1 (h_1\rightarrow
xx)}{\Gamma_1}$ where $\Gamma_1 (h_1\rightarrow xx)= \cos^2\alpha \Gamma^{\rm
SM}(h\rightarrow xx)$ is the decay width of $h_1$ decaying into $xx$ final
state while $\Gamma_1$ is the total decay width of $h_1$ in the present
model. Hence, the signal strength of $h_1$ in Eq.~\ref{4} can be written in the
form
\bea 
R_1 = c_{\alpha}^4 \frac{\Gamma^{\rm SM}}{\Gamma_1} \,\, ,
\label{5}
\eea 
where we have denoted $\cos\alpha$ as $c_{\alpha}$. It is to be noted that apart
from the decay into SM particles the SM-like scalar $h_1$ can also have
invisible decay mode into dark matter particles. Therefore the total decay
width of $h_1$, in the present model, can be written as
% $\Gamma^{\rm SM}_1$ is total decay
%width of
%SM Higgs of mass $m_1=125$ GeV and $\Gamma_1$ is the total decay width of 
%SM-like scalar $h_1$ in our model,
\bea
\Gamma_1= c_{\alpha}^2 \Gamma^{\rm SM} + \Gamma_1^{\rm inv} \,\, .
\label{6a}
\eea  
In Eq.~\ref{6a}, $\Gamma_1^{\rm inv}$ is the invisible decay
width $h_1$ for the channel $h_1\rightarrow f_1 \bar{f_1}$.
For $m_1 > 2 \,m$ the expression of invisible decay width
of $h_1$ is given by
\bea
\Gamma_1^{\rm inv}=\frac{m_1}{8\pi}\frac{m^2}{v_s^2}s_{\alpha}^2
\left(1-\frac{4m^2}{m_1^2}\right)^{3/2}\,\, ,
\label{7}
\eea
since coupling between $h_1$ and dark matter candidate is proportional to
$\frac{m}{v_s}s_{\alpha}$. In the
above, $v_s$ is the VEV of SU(2)$_{\rm H}$ Higgs doublet $\rm \Phi$ and 
$s_{\alpha}=\sin\alpha$. Similarly for the other scalar involved in our model,
the signal strength $R_2$ is expressed as
\bea
R_2 =\frac {\sigma (pp\rightarrow h_2)} {\sigma^{\rm SM}(pp\rightarrow h)}
\frac {{\rm Br} (h_2\rightarrow xx)} {{\rm Br}^{\rm SM}(h\rightarrow xx)} \,\,  
\label{8}
\eea  
with $\sigma(pp\rightarrow h_2)$ being the production cross-section of $h_2$ and 
${\rm Br}(h_2\rightarrow  x x)$ is decay branching ratio of $h_2$ to any final
state. However in this case, the Standard Model
predictions $\sigma^{\rm SM}(pp\rightarrow h)$
and ${\rm Br}^{\rm SM} (h\rightarrow xx)$ are
computed for the mass of SM Higgs boson $m_h = m_2$.
Using the similar approach we used to calculate $R_1$ and replacing
$h_1,\cos \alpha$ etc. by $h_2,\sin \alpha$ the signal strength $R_2$
of $h_2$ can be expressed as
\bea 
R_2 = s_{\alpha}^4 \frac{\Gamma^{\rm SM}(m_h=m_2)}{\Gamma_2} \,\, ,
\label{9}
\eea 
where $\Gamma^{\rm SM}(m_h=m_2)$ is the total decay width
of SM Higgs boson if it has mass $m_h = m_2$
while $\Gamma_2$ is the total decay width for the non-SM scalar boson
$h_2$
\bea
\Gamma_2= s_{\alpha}^2 \Gamma^{\rm SM} (m_h=m_2) + \Gamma_2^{\rm inv} \,\, .
\label{6}
\eea  
The coupling between dark matter and $h_2$ depends on the factor
$\frac{m}{v_s}\cos_{\alpha}$. Hence, invisible decay width of $h_2$
($\Gamma_2^{\rm inv}$) for $m_2>2m$ is given as 
\bea
\Gamma_2^{\rm inv}=\frac{m_2}{8\pi}\frac{m^2}{v_s^2}c_{\alpha}^2
\left(1-\frac{4m^2}{m_2^2}\right)^{3/2}\,\, .
\label{11}
\eea 
%In this work, we consider two cases of hidden scalar (non-SM scalar) mass $m_2$
%such that (a) $m_1>m_2=100$ GeV and (b) $m_1<m_2=200$ GeV in the framework of
%our proposed model.
As stated earlier, we consider $h_1$ with mass $m_1=125$ GeV to be the Higgs
like scalar and infer $R_1>0.8$ \cite{ATLAS:2012xmd} and invisible decay 
branching ratio ${\rm Br}^1_{\rm inv}\leq 0.2$ \cite{Belanger:2013kya} where 
${\rm Br}^1_{\rm inv}=\Gamma^1_{\rm inv}/\Gamma^1$ is defined as the ratio of
invisible decay width to the total decay width.  
\item {\bf Dark matter relic density} -
\begin{figure}[h!]
\centering
\includegraphics[height=7 cm, width=10 cm,angle=0]{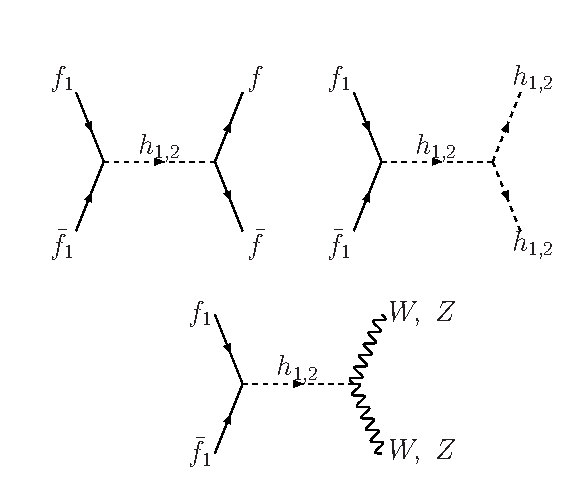}
\caption{Feynman diagrams for dark matter annihilation into fermions (quarks and leptons),
gauge bosons and scalars contributing to DM annihilation cross-section.}
\label{feyn}
\end{figure}
The DM relic density as measured by PLANCK satellite experiment is given as
\cite{Ade:2013zuv}
\be
\Omega_{\rm DM} h^2 = 0.1199{\pm 0.0027}\,\, .
\label{12}
\ee 
In Eq.~\ref{12}, $h$ is the Hubble parameter measured in the unit of
100 km~s$^{-1}$~Mpc$^{-1}$. We calculate the relic density for the fermionic
(SU(2)$_{\rm H}$) dark matter candidate in the assumed dark sector in our model
by solving the Boltzmann equation. Relic density of the DM candidate is obtained
by solving the Boltzmann equation \cite{kolb}
\bea
\frac{{\rm d}n}{{\rm d} t} + 3 {\rm H}n &=& - \langle \sigma {\rm v} \rangle
(n^{2}-n_{\rm{eq}}^{2})\,\, ,
\label{13}
\eea    
where $n$ is the number density of DM particle and $n_{\rm eq}$ is the same in
equilibrium. In Eq.~\ref{13}, $\langle \sigma {\rm v} \rangle$ is the 
thermal averaged annihilation cross-section of DM particle into SM sector
and $\rm H$ is Hubble parameter. Solution to Eq.~\ref{13} gives the DM
relic abundance of the form
\bea
\Omega_{\rm{DM}}{\rm h}^2 &=& \frac{1.07\times 10^9 x_f}
{\sqrt{g_*}M_{\rm Pl}\langle \sigma {\rm v} \rangle}\,\, ,
\label{14}
\eea  
where $g_*$ is the effective number of d.o.f (degrees of freedom), 
$M_{\rm Pl}$ is the PLANCK mass ($\sim 1.22\times 10^{19}$ GeV) and 
$x_f=m/T_f$ with $T_f$ being the freeze out temperature of the DM species
respectively. We compute the freeze out temperature $T_f$ by solving 
iteratively the following equation
\bea
x_f &=& \ln \left ( \frac{m}{2\pi^3}\sqrt{\frac{45M_{\rm{Pl}}^2}{2g_*x_f}}
\langle \sigma {\rm v} \rangle \right )\,\, . 
\label{15}
\eea
In order to obtain the freeze out temperature of DM and hence its relic density
using Eqs. \ref{14}, \ref{15} we need to calculate the thermal average of
the product between total DM annihilation cross-section ($\sigma$)
and the relative velocity (${\rm v}$) of two annihilating DM particles.
The expression for the thermally averaged DM annihilation cross-section
into all possible final states is given as
\bea
\langle \sigma {\rm v} \rangle=\frac{1}{8m^4\,T_f\,K_2^2(m/T_f)}
\int_{4m^2}^\infty ds~\sigma (s)~(s-4m^2)\,\sqrt{s}\,
K_1(\sqrt{s}/T_f) ,    
\label{16}
\eea
where the factors $K_i,~(i=1,2)$ are the modified Bessel functions and
$\sqrt{s}$ being the centre of mass energy. In the present formalism
dark matter candidate $f_1$ can annihilate into the SM particles
through $s$-channel processes mediated by the scalar bosons $h_1$ and $h_2$.
In the above Eq. \ref{16}, $\sigma (s)$ denotes the total annihilation
cross-section of dark matter into all possible final states which are
allowed by the Lagrangian given in Eq. \ref{prob2}. Feynman diagrams
for different annihilation channels of $f_1$ are shown in Fig.~\ref{feyn}.
The expressions of $\sigma {\rm v}$ for different final state annihilation of
dark matter into SM particles are derived from the Feynmann diagrams shown in
Fig.~\ref{feyn}. The value of $\sigma {\rm v}$ obtained for DM
annihilation into SM fermion and antifermion pairs ($f\bar{f}$) at the final
state is of the form

{\bea
\sigma {\rm v}_{f \bar f} &=& N_c \frac{m^2}{v_s^2}\frac{s_{\alpha}^2c_{\alpha}^2}{8\pi}
\frac{m_f^2}{v^2}\left(1-\frac{4m_f^2}{s}\right)^{3/2}
(s-4m^2) \left [\frac{1}{(s-m_1^2)^2+m_1^2\Gamma_1^2}+
\frac{1}{(s-m_2^2)^2+m_2^2\Gamma_2^2} \right . \nonumber \\ 
&& \left . -\frac{2(s-m_1^2)(s-m_2^2)+2m_1m_2\Gamma_1\Gamma_2}
{[(s-m_1^2)^2+m_1^2\Gamma_1^2][(s-m_1^2)^2+m_1^2\Gamma_2^2]} \right ]\,\, ,
\label{17}
\eea}
where $m$ is the DM mass and $m_f$ is the mass of specific fermion ($f=$ quark
or lepton). In Eq.~\ref{17} $v$ and $v_s$ are the vacuum expectation values of
SM Higgs doublet and dark Higgs doublet, $N_c$ is the colour quantum number (3
for quarks and 1 for leptons). Moreover, $\Gamma_1$, $\Gamma_2$ are the
total decay widths of the scalar bosons $h_1$, $h_2$ and the expressions
of  $\Gamma_1$ and $\Gamma_2$ are given in Eqs. \ref{6a}, \ref{6}.
%Term $s$ in Eq.~\ref{17} is the square of centre of
%momentum energy of the $s$-channel process.
We also calculate $\sigma {\rm v}$ for
$W^+W^-$ and $ZZ$ channels which proceed through the $s$-channel exchange of
scalar bosons $h_1$, $h_2$ (see Fig. \ref{feyn}). The expressions
of $\sigma {\rm v}_{W^+ W^-}$ and $\sigma {\rm v}_{ZZ}$ are furnished below
{\small
\bea
\sigma {\rm v}_{W^+W^-} &=& \frac{m^2}{v_s^2}\frac{s_{\alpha}^2c_{\alpha}^2}{8\pi s}
\left(1-\frac{4m_W^2}{s}\right)^{1/2} \left(\frac{2m_W^2}{v}\right)^2
\left(1+\frac{(s/2-m_W^2)^2}{2m_W^4}\right)
(s-4m^2) \left [\frac{1}{(s-m_1^2)^2+m_1^2\Gamma_1^2} \right . \nonumber \\
&& \left . +\frac{1}{(s-m_2^2)^2+m_2^2\Gamma_2^2}  
-\frac{2(s-m_1^2)(s-m_2^2)+2m_1m_2\Gamma_1\Gamma_2}
{[(s-m_1^2)^2+m_1^2\Gamma_1^2][(s-m_1^2)^2+m_1^2\Gamma_2^2]} \right ]\,\, ,
\label{18}
\eea}
and
{\small
\bea 
\sigma {\rm v}_{ZZ} &=& \frac{m^2}{v_s^2}\frac{s_{\alpha}^2c_{\alpha}^2}{16\pi s}
\left(1-\frac{4m_Z^2}{s}\right)^{1/2} \left(\frac{2m_Z^2}{v}\right)^2
\left(1+\frac{(s/2-m_Z^2)^2}{2m_Z^4}\right)
(s-4m^2) \left [\frac{1}{(s-m_1^2)^2+m_1^2\Gamma_1^2} \right . \nonumber \\
&& \left . +\frac{1}{(s-m_2^2)^2+m_2^2\Gamma_2^2}  
-\frac{2(s-m_1^2)(s-m_2^2)+2m_1m_2\Gamma_1\Gamma_2}
{[(s-m_1^2)^2+m_1^2\Gamma_1^2][(s-m_1^2)^2+m_1^2\Gamma_2^2]} \right ]\,\, .
\label{19}
\eea}
In the above, $m_W$ and $m_Z$ denotes the respective masses of $W$ and $Z$
bosons. Annihilations of DM particles into scalar bosons $h_1$ and $h_2$ are also
taken into account. The process of DM annihilation into scalars
$h_1$ or $h_2$ is also scalar mediated, depends on scalar
couplings between $h_1$ and $h_2$. The $s$-channel annihilation cross-section of
$f_1$ annihilating into the pairs of $h_1$ and $h_2$, calculated using $f_1
\bar{f_1} \rightarrow h_i h_i,~i=1,2$ annihilation diagram, takes the following
form
{\small
\bea
\sigma {\rm v}_{h_1h_1} &=& \frac{1}{16\pi s}\frac{m^2}{v_s^2}
\left(1-\frac{4m_1^2}{s}+\frac{4m_1^2(m_1^2-1)}{s^2}\right)^{1/2}
(s-4m^2) \left [\frac{s_{\alpha}^2\lambda_{111}^2}{(s-m_1^2)^2+m_1^2\Gamma_1^2} 
\right . \nonumber \\&& \left .
+\frac{c_{\alpha}^2\lambda_{211}^2}{(s-m_2^2)^2+m_2^2\Gamma_2^2}  
-\frac{2s_{\alpha}c_{\alpha}\lambda_{111}\lambda_{211}((s-m_1^2)
(s-m_2^2)+2m_1m_2\Gamma_1\Gamma_2)}
{[(s-m_1^2)^2+m_1^2\Gamma_1^2][(s-m_1^2)^2+m_1^2\Gamma_2^2]} \right ]\,\, ,
\label{20}
\eea}
and
{\small
\bea
\sigma {\rm v}_{h_2h_2} &=& \frac{1}{16\pi s}\frac{m^2}{v_s^2}
\left(1-\frac{4m_2^2}{s}+\frac{4m_2^2(m_2^2-1)}{s^2}\right)^{1/2}
(s-4m^2) \left [\frac{s_{\alpha}^2\lambda_{122}^2}{(s-m_1^2)^2+m_1^2\Gamma_1^2}
\right . \nonumber \\&& \left .
+\frac{c_{\alpha}^2\lambda_{222}^2}{(s-m_2^2)^2+m_2^2\Gamma_2^2}  
-\frac{2s_{\alpha}c_{\alpha}\lambda_{122}\lambda_{222}((s-m_1^2)
(s-m_2^2)+2m_1m_2\Gamma_1\Gamma_2)}
{[(s-m_1^2)^2+m_1^2\Gamma_1^2][(s-m_1^2)^2+m_1^2\Gamma_2^2]} \right ]\,\, ,
\label{21}
\eea}
where, $\lambda_{ijk}$ is the coupling for the vertex involving three scalar
fields $h_i h_j h_k$. The expressions for the scalar
couplings $\lambda_{111},~\lambda_{211},~\lambda_{122}$
and $\lambda_{222}$ are given in Appendix A. 
We calculate the thermally averaged annihilation cross-section
of the present DM candidate using Eqs.~\ref{16}-\ref{21}
We then compute the freeze out temperature $T_f$ by solving Eq.~\ref{15} and
finally obtain the relic density of $f_1$
at the present epoch from Eq.~\ref{14}.
%From Eqs.~\ref{17}-\ref{21} it can be seen that the annihilation
%cross-sections derived for specific final states proceed via $s$-channel.
\item {\bf DM Direct Detection} -
Direct detection of DM particle is based on the scattering of the DM particle
with the target nucleus of the detector material. Fermionic dark matter in
the present model can undergo elastic scattering with the detector nucleus.
This elastic scattering of the DM and the nucleus will transfer a recoil energy
to the target nucleus which is then calibrated. From the non-observance of such
elastic scattering events the direct detection experiments give the upper bound
of elastic scattering cross-sections for different possible masses of dark
matter. The scattering cross-section is expressed as cross-section per nucleon
for enabling direct comparison of the results from different experiments. 
In the present model DM fermion of mass $m$ can interact with the target
nucleus through t-channel Higgs mediated processes through both $h_1$ and $h_2$.
The spin-independent (SI) elastic scattering cross-section off the detector 
material normalised to per nucleon can be written as \cite{LopezHonorez:2012kv}
\be
\sigma_{\rm {SI}} = \frac{\sin^22\alpha}{4\pi}\frac{m^2}{v_s^2}m_r^2
\left (\frac{1}{m_1^2}-\frac{1}{m_2^2} \right )^2 \lambda_p^2 
\label{22}
\ee   
where $m_r=\frac{mm_p}{m+m_p}$ is the reduced mass for the DM-nucleon system
and $\lambda_p$ \cite{LopezHonorez:2012kv} is given in terms of the form
factors $f_q$, proton mass $m_p$ as
\be
\lambda_p=\frac{m_p}{v}\left[\sum_q f_q+ \frac{2}{9}\left(1-\sum_q
f_q\right)\right] \simeq1.3\times10^{-3} \,\, .
\label{23}
\ee
Using Eqs.~\ref{22}-\ref{23}, we calculate the spin independent elastic
scattering cross-section of the DM fermion off the nucleon and compare it with
the experimental bounds from LUX \cite{Akerib:2013tjd}.

Note that both DM annihilation cross-section and DM-nucleon scattering
cross-section depend on an effective coupling 
$g_{eff} = |\frac{m}{v_s}s_{\alpha}c_{\alpha}|$ (Eqs.~\ref{17}-\ref{22}). This
effective coupling is a useful parameters to explain the dark matter
phenomenology in the present framework. Further discussions on the effective
coupling are given later in Sec.~\ref{S:res}.
 
\item {\bf DM Indirect Detection}
The existence of DM has now been well established from gravitational evidences
in astrophysical scale. Indirect search of DM focuses on the non-gravitational
search of DM candidate and explores the particle physics nature of DM. The
astrophysical sites such as Galactic Centre (GC), dwarf galaxies etc. are of
great interest since dark matter can be trapped and accumulate at GC due to the
enormous gravity in the region of GC and the mass to luminosity ratio of dwarf
galaxies indicate the presence of dark matter in large magnitude.  
These sites are suitable for indirect search of DM as DM particles trapped in
these regions can undergo annihilation into various SM particles which can
further produce gamma rays, neutrinos etc. Thus any observed excess in the
fluxes of $\gamma$-ray, positron, anti-proton from such sites can indicate
DM annihilation processes in those sites if other astrophysical phenomena cannot
explain the observed excess.
Fermi-LAT \cite{Ackermann:2012qk} searches for the
excess emission of $\gamma$-rays originating from GC and dwarf galaxies. Observation
of the excess in $e^+/e^-$ and $p/ \bar p$ flux is performed by AMS-02
\cite{Aguilar:2013qda} experiment. In this Section we will study Fermi-LAT
observed gamma ray flux results from the centre of Milky Way and surrounding dwarf
spheroidal galaxies (dSphs).

The expression for the differential $\gamma$-ray flux obtained from a region of 
interest (ROI) subtends a solid angle $\rm d\Omega$ centered at GC is given
as
\bea
\frac{\rm d\Phi}{\rm dE d\Omega}=\frac{1}{8\pi m_{DM}^2}J
\sum_f{\langle \sigma {\rm v}\rangle}_f\frac{dN_f}{dE_{\gamma}} \,\, ,
\label{24}
\eea
where ${\langle \sigma {\rm v}\rangle}_f$ is the average thermal annihilation 
cross-section of DM particles annihilating into final state particle $f$
and $\frac{dN_f}{dE_{\gamma}}$ is the photon energy spectrum of DM annihilation
into the same. The factor $J$ appearing in Eq.~\ref{24} is related to the
quantity of dark matter present at the astrophysical site considered and is
expressed in terms of dark matter density as 
\bea
J=\int_{\rm los}\rho^2(r(s,\theta))ds\,\,  .
\label{25}
\eea 
In Eq.~\ref{25} the line of sight (los) integral is performed over an angle
$\theta$ , is the angular aperture between the line connecting GC to the Earth
and the direction of line of sight. In the above Eq.~\ref{25}, 
$r=\sqrt{r_{\odot}^2+s^2-2r_{\odot}s\cos\theta}$ where $r_{\odot}=8.5$ kpc,
is the distance to the Sun from GC. It is clear from the expression of 
Eq.~\ref{25} that value of $J$ factor is dependent on the nature of the chosen 
$\rho(r)$ factor i.e, DM halo density profile $\rho(r)$. In the present work, we
consider Navarro-Frenk-White (NFW) \cite{Navarro:1995iw} halo profile. DM
density distribution for the NFW halo profile is given as
\bea
\rho(r)=\rho_0\frac{(r/r_s)^{-\gamma}}{(1+r/r_s)^{3-\gamma}}\, .
\label{26}
\eea 
where $r_s=20$ kpc is the characteristic distance and $\rho_0$ is normalised to
local DM density i.e., $\rho_{\odot}=0.4~{\rm{GeV~cm^{-3}}}$ at a distance 
$r_{\odot}$ from GC.

The analysis by Daylan et. al. \cite{Daylan:2014rsa} of Fermi-LAT data suggests
an excess in $\gamma$-ray in the $\gamma$ energy range of 2-3 GeV at GC. The
same analysis demonstrates that this excess can be explained by the annihilation
of 31-40 GeV DM into $b \bar b$ with $ {\langle \sigma {\rm v} \rangle}_{b \bar
b} = 1.4-2.0 \times 10^{-26}~{\rm cm}^3{\rm s}^{-1}$. 
In this work \cite{Daylan:2014rsa}, inner galaxy gamma ray flux  ($5^0$ from GC)
is calibrated using NFW halo profile with $\gamma=1.26$ and local DM density
$\rho_{\odot}= 0.3~{\rm{GeV~cm^{-3}}}$. In a recent work by Calore, Cholis and
Weniger (CCW) \cite{Calore:2014xka} detailed analysis is performed for the
GC $\gamma$-rays along with the systematic uncertainties using 60 galactic
diffusion excess (GDE) models.
Results from CCW analysis provides a best fit for
DM annihilation into $b \bar b$ having mass $49^{+6.4}_{-5.4}$  GeV with  
$ {\langle \sigma {\rm v} \rangle}_{b \bar b} =
1.76^{+0.28}_{-0.27} \times 10^{-26}~{\rm cm}^3{\rm s}^{-1}$. However, CCW 
analysis of Galactic Centre excess (GCE) for gamma ray have also considered
generalised NFW profile ($\gamma$=1.2, $\rho_{\odot}=0.4~{\rm{GeV~cm^{-3}}}$)
for a different region of  interest (ROI) with galactic latitude $|l|\leq20^0$
and longitude $|b|\leq20^0$ masking out inner $|b|\leq2^0$. In another work 
P. Agrawal et. al. \cite{Agrawal:2014oha} reported that annihilation of
heavier dark matter (upto 165 GeV for $b \bar b$ channel) can also explain
the observed GCE in $\gamma$-ray when uncertainties in DM halo profile (NFW)
and the $J$-factor are taken into account. However in the
present work, we do not consider any such uncertainties in halo profiles or 
$J$ vaules and use the canonical NFW halo profile used in CCW analysis. Using
Eqs.~\ref{24}-\ref{26}, we calculate the $\gamma$-ray flux (in GeV cm$^{-2}$
s sr$^{-1}$) for the ROI described in CCW analysis for Fermi-LAT data.
As mentioned earlier we consider for our calculations the
NFW profile with $\gamma=1.2$ and $\rho_{\odot}=0.4~{\rm{GeV~cm^{-3}}}$.

Apart from the GC region, dwarf galaxies of the
Milky-Way galaxy are also of great significance for indirect search of DM
as these galaxies are supposed to be rich in dark matter.
Recent analyses of $\gamma$-ray fluxes from 15 Milky-Way dSphs
reported by Fermi-LAT \cite{Ackermann:2015zua} provide a limit on DM mass and 
corresponding thermally averaged annihilation cross-section 
${\langle \sigma {\rm v} \rangle}_f$ into different channels  
$f~(\tau~{\rm and}~b)$. Fermi-LAT have used their 6 year data collected by
Fermi Large area Telescope and performed an analysis for 15 dSphs using ``pass-8
event level analysis`` (see \cite{Ackermann:2015zua} and references therein). 
In an another work \cite{Drlica-Wagner:2015xua} Fermi-LAT in collaboration along
with Dark Energy Survey (DES) collaboration also provide similar bound on
${\langle \sigma {\rm v} \rangle}_f$ where they include data for 8 new dSphs.
For both the analysis presented in
\cite{Ackermann:2015zua,Drlica-Wagner:2015xua} a canonical
NFW halo profile ($\gamma=1$) is considered, and the astrophysical $J$
factors are measured over a solid angle $\Delta\Omega=2.4\times10^{-3}~{\rm sr}$
with angular radius $0.5^0$. Independent searches carried out by
Fermi-LAT \cite{Ackermann:2015zua} and DES-Fermi-LAT collaboration on 15 
previously discovered and 8 recently discovered
different dSphs reported no significant excess in observed $\gamma$-ray.
Results from the DES dSphs \cite{Drlica-Wagner:2015xua} also predicts an upper
bound to the observed $\gamma$-ray energy flux with 95\% confidence limit (C.L.)
for 8 newly found dSphs. Gamma ray flux for dwarf galaxies when integrated for
an energy range extending over a region of solid angle $\Delta\Omega$ is
expressed as
\bea
\Phi=\frac{{\langle \sigma {\rm v}\rangle}}{8\pi m_{DM}^2}J
\int_{E_{\rm min}}^{E_{\rm max}}\frac{dN}{dE_{\gamma}}dE_{\gamma} \,\, ,
\label{27}
\eea
where $\frac{dN}{dE_{\gamma}}$ is the $\gamma$-ray.
The expression of flux presented in Eq.~\ref{27}
is calculated for a single final state annihilation of DM. Hence, summation over
different final channels is not needed. Form of $J$ factor appearing in 
Eq.~\ref{27} is different from Eq.~\ref{25} and written as
\bea
J=\int_{\Delta\Omega}\int_{\rm los}\rho^2(r(s,\theta))ds\, ,
\label{28}
\eea 
calculated over a solid angle $\Delta\Omega=2.4\times10^{-3}$ sr subtended by
the ROI ($0.5^0$ angular radius) for NFW halo profile ($\gamma=1$). The density
distribution function for NFW profile with $\gamma=1$ is then 
\bea
\rho(r)=\rho_0\frac{r_s^3}{r(r_s+r)^2}\, ,
\label{29}
\eea   
where $r_s$ is the NFW scale radius and $\rho_0$ represents the characteristic
density for the dSphs. In the case of Fermi-LAT analysis, $J$ factors for
different dSphs are adopted from Ref. \cite{Ackermann:2015zua}. We use values 
of $J$ factor from \cite{Drlica-Wagner:2015xua} for computing
gamma ray flux for 8 DES dSphs for the dark matter candidates in our model.
However, it is to be noted that
$J$ factors for DES dSphs candidates are obtained assuming the point like
dSphs instead of having spatial extension (as in the case of
\cite{Ackermann:2015zua}) to avoid the uncertainties in halo profile arising
from spatial extension. Calculation of gamma ray flux is also based on the
assumption that the spectrum $\frac{dN}{dE_{\gamma}}$ follows the conventional
power law $\frac{dN}{dE_{\gamma}}\sim\frac{1}{E^2}$.
As mentioned earlier, study of 15 dSphs by Fermi-LAT and 8 other dSphs
by DES-Fermi-LAT collaboration found no significant excess in $\gamma$-ray 
from these dwarf galaxies. 
However, a recent search on a newly discovered dwarf
galaxy Reticulum 2 (Ret2) in a work by Geringer-Sameth et. al 
\cite{Geringer-Sameth:2015lua} has reported an excess in observed $\gamma$-ray signal.
In the present work, we calculate the $\gamma$-ray flux for 
annihilation of hidden SU(2)$_{\rm H}$ fermionic dark matter into $\gamma$-ray
through different SM final states and explore whether the model can account for 
GCE in $\gamma$-ray and also satisfies the bounds on gamma ray flux from dwarf
satellite galaxies. 

As mentioned earlier, in the present model dark matter candidate ($f_1$) is
fermionic in nature and it interacts with the visible world (SM particles)
through the exchange of two real scalar bosons $h_1$ and $h_2$. As a result the
annihilation cross-sections of the DM dark candidate $f_1$ into the final states
that composed of SM particles (mainly light quarks and leptons) are proportional
to the square of relative velocity (${\rm v}^2$) between the annihilating 
dark matter particles (p wave process). Now the averaged DM relative velocity
is proportional to $\sim \sqrt{3/x}$ \cite{Ibe:2008ye}-\cite{Guo:2009aj} with
$x=\frac{m}{T}$ is a dimensionless quantity and $T$ being the temperature of
the Universe. Hence, in our model, the thermally averaged annihilation
cross-section used for computing DM relic density, at $x\sim 20-30$, is
different from the annihilation cross-section (for $x \sim 3 \times 10^6$
\cite{Ibe:2008ye}-\cite{Guo:2009aj}) needed to calculate $\gamma$-ray flux at
the Galactic Centre and dwarf galaxies. The latter quantity is velocity
suppressed as the average DM relative velocity is $\sim 10^{-3}$ when the
annihilation of DM occurs at the GC. Among all the annihilation channels of
$f_1$, the annihilation mode $f_1 \bar{f_1} \rightarrow b \bar{b}$ plays a
significant role for the $\gamma$-ray excess observed from GC and dwarfs
satellite galaxies as it is the most dominant annihilation channel for the
considered mass range of DM. In order to explain the GC gamma-excess by DM
annihilation to $b\bar{b}$, the annihilation cross-section should be $\sim
1.76^{+0.28}_{-0.27}\times 10^{-26}$ cm$^3/$s \cite{Calore:2014xka}. Although in
the present case, the thermally averaged annihilation
cross-section for the $b\bar{b}$ annihilation is quite small, however the
quantity $\langle {\sigma {\rm v}}\rangle_{b\bar{b}}$ can be significantly
enhanced using Breit-Wigner resonant enhancement mechanism
\cite{Ibe:2008ye}-\cite{Guo:2009aj}. Breit-Wigner enhancement occurs only when
the mass of the dark matter ($m$) is nearly
equal to half of the mediator mass (in our case it is the mass of $h_2$).
Therefore, we have defined the mass of the hidden sector scalar boson ($h_2$)
and the centre of mass energy $\sqrt{s}$ in the following way
\begin{eqnarray}
m^2_2 = 4m^2(1-\delta)\,\,{\rm and}\,\, s=4m^2(1+z)\,\,,
\label{30} 
\end{eqnarray}
where $\delta<0$ represents the physical pole and $z$ is the measure of excess
centre of momentum energy scaled by $4m^2$. In terms of $z$, Eq.~\ref{16} for
the $b\bar{b}$ annihilation channel, can now be written as
\begin{eqnarray}
{\langle{\sigma {\rm v}}\rangle}_{b\bar{b}} =
\frac{4\,x}{K^2_2(x)}\int^{z_{eff}}_0 dz\,\sigma(z)_{b\bar{b}}\,z
\,\sqrt{1+z}\,K_1(2\,x\,\sqrt{1+z})
\label{31}
\end{eqnarray}
with the expression of $\sigma(z)_{b\bar{b}}$ is given by
\footnote{Since the Breit-Wigner enhancement occurs when $m\sim m_2/2$, as a
result only the term proportional to $\frac{1}{(s-m_2^2)^2+m_2^2\Gamma_2^2}$
will dominanatly contribute to the annihilation cross-section appearing in
Eq.~\ref{17}.}
\begin{eqnarray}
\sigma(z)_{b\bar{b}} = \frac{g_{c}}{4 m^2}\,
\frac{\sqrt{z}}{1+z} \,
\frac{\left(1+z-\frac{m^2_b}{m^2}\right)^{3/2}}
{\left[(z+\delta)^2+\gamma^2_2 (1-\delta)^2\right]}\,\,,
\label{32}
\end{eqnarray}
and
\begin{eqnarray}
g_{c} = \frac{N_c}{16\,\pi}\,
\left(\frac{m \cos \alpha}{v_s} \frac{m_b \sin \alpha}{v} \right)^2 \,
\label{33}
\end{eqnarray}
where $\gamma_2 = \frac{\Gamma_2}{m_2}$, $\Gamma_2$ being the total decay
width of $h_2$ of mass $m_2$. It is to be noted that the upper limit of
the above integration should be $\infty$ (see Eq. \ref{16}), however the
integrand becomes negligibly small when $z$ approaches to $z_{eff} \sim {\rm
max}[4/x,\,2|\delta|]$ for $\delta<0$ \cite{Basak:2014sza},\cite{Guo:2009aj}.
Using the above prescription, we calculate the thermally averaged
annihilation cross-section $\langle{\sigma{\rm v}}\rangle_{b\bar{b}}$ of the dark
matter candidate $f_1$ for GC and dwarf spheroidal galaxies. The actual values of
$\langle{\sigma{\rm v}}\rangle_{b\bar{b}}$, $\gamma_2$ and $\delta$ for the two
chosen bench mark points (BP1, BP2) are given in Table~\ref{tab1} of
Sec.~\ref{S:res}. We have found that for $|\delta|\sim10^{-3}$ the annihilation
cross-section $\langle{\sigma{\rm v}}\rangle_{b\bar{b}} \sim 1.9\times 10^{-26}$
cm$^3/$s which can explain the excess of gamma ray flux in GC\footnote{Similar
results for Breit-Wigner enhancement of dark matter annihilatin cross-section
have been reported in \cite{Basak:2014sza}.}.

\end{itemize}
\section{Calculational procedures and Results}
\label{S:res}
In this section we present the computation of dark matter annihilation
cross-sections as also the DM-nucleon elastic scattering cross-sections. They
are required for the calculation of relic densities and the comparison of the
latest DM scattering cross-section bound given by the LUX direct detection
experiment. The invisible decay widths and signal strengths for the SM-like
scalar is also calculated in order to constrain the model parameter space.
The gamma ray flux are then computed within the framework of SU(2)$_{\rm H}$
fermionic dark matter for galactic centre as also for dwarf galaxies and the
results are compared with the experimental analysis.

\subsection{Constraining the model parameter space}
\label{SS:direct}

The fermionic dark matter in the present model
can  annihilate through scalar mediated ($h_1$ and $h_2$) $s$-channel processes.
As mentioned in Sec.~\ref{S:cons}, the model parameter space is first
constrained by the vacuum stability conditions given in Eq.~\ref{2}. The signal
strengths $R_1$ and $R_2$ for the Higgs doublets $h_1$ (SM) and $h_2$ (dark
sector) are then computed using Eqs.~\ref{12}-\ref{16}. With the chosen
constraints on $R_1$
($R_1\ge$ 0.8, Ref.~\cite{ATLAS:2012xmd}) the invisible decay branching ratio of
SM-like Higgs ${\rm Br}^1_{\rm inv}$ is calculated and the parameter space is
further constrained by LHC experiment limit of ${\rm Br}^1_{\rm inv}$ (${\rm
Br}^1_{\rm inv}\le~0.2$ \cite{Belanger:2012zr}). The parameter space thus
constrained is then used to
compute the thermal averaged annihilation cross-section ${\langle \sigma {\rm v}
\rangle}$ of the present fermionic dark matter candidate and the relic density
is obtained by solving the Boltzmann equation (using Eqs.~\ref{13}-
\ref{16}). The annihilation cross-sections are computed with the calculated
analytical formulae given in Eqs.~\ref{17}-\ref{21} with two choices of VEV for 
$\Phi$ (dark Higgs doublet) namely $v_s=$ 246 GeV and 500 GeV. In our
calculation we consider the mass $m_1$ of the SM-like Higgs boson $h_1$ to be
125 GeV. The calculation is performed for two values of the dark sector scalar
$h_2$ masses and they are $m_2=$ 100 GeV and 110 GeV. These relic densities are
compared with the dark matter relic density given by PLANCK \cite{Ade:2013zuv}.
Thus PLANCK result further constrains the parameter space of our model. With
this available parameter space we evaluate the dark matter-nucleon spin
independent scattering cross-section ($\sigma_{\rm SI}$) for the purpose of
comparing our results with those given by the dark matter direct detection
experiments such as LUX, XENON100 etc. In this way we restrict our model
parameter space by different experimental results.

\begin{figure}[h!]
\centering
\subfigure[]{
\includegraphics[height=7 cm, width=7 cm,angle=0]{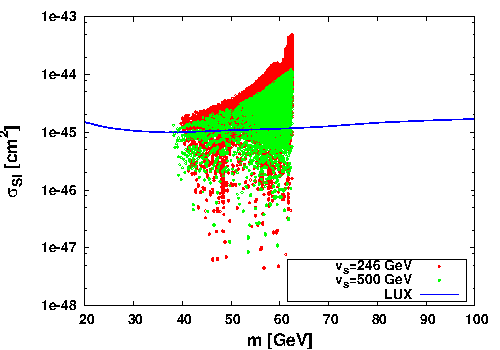}}
\subfigure []{
\includegraphics[height=7 cm, width=7 cm,angle=0]{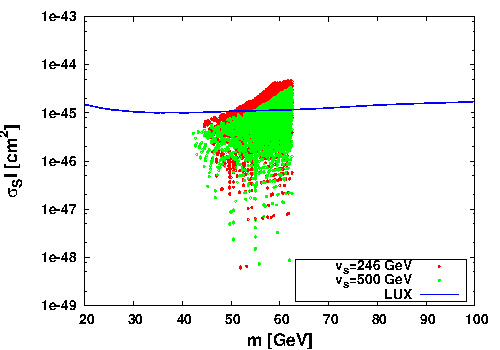}}
\caption{The allowed range of $m-\sigma_{\rm SI}$ parameter space obtained for
$m_2=$ 100 GeV (left panel) and $m_2=$ 110 GeV (right panel) plotted using the
bounds from vacuum stability, LHC constraints on SM Higgs and relic abundance
of DM obtained from PLANCK \cite{Ade:2013zuv}. Limits on DM-nucleon 
cross-section from LUX \cite{Akerib:2013tjd} is also plotted (blue line) for
comparison.}
 \label{fig1}
\end{figure}

In Fig.~\ref{fig1}a and  Fig.~\ref{fig1}b we show the calculated values of
$\sigma_{\rm SI}$ with different DM mass in the present model where the
conditions from vacuum stability, bound on SM Higgs signal strength and DM relic
density results from PLANCK have been imposed. We first choose certain
values of $m_1$ and $m_2$ and vary the couplings $\lambda_i,~i=1$ to $3$
(satisfying vacuum stability conditions given in Eq.~\ref{2}) for two different
values of $v_s$ which also constrain the mixing angle $\alpha$ through the 
Eq.~\ref{prob7}. Here we want to mention that we have varied $\lambda_1$ 
and $\lambda_2$ in the range 0 to 0.2 with the values of both $\lambda$'s are
evenly spread within the considered range. Consequently the value of the 
parameter $\lambda_3$ becomes fixed by the vacuum stability criteria given in
Eq.~\ref{2} which is also varied with equal interval in the range 
$|\lambda_3|<2\sqrt{\lambda_1 \lambda_2}$. The
model parameter space thus obtained is then
further constrained by imposing the conditions $R_1>0.8$ and
$Br^1_{\rm inv}<0.2$ from LHC results. Using this restricted model
parameter space satisfying both vacuum stability and LHC bounds,
we therefore calculate the relic density of the dark matter
candidate $f_1$ by solving the Boltzmann equation (Eq. \ref{13})
for different values of DM mass. Finally, we consider
specific range of model parameter space which is in agreement
with DM relic density reported by PLANCK experiment
and for these parameter space we compute the spin-independent
direct detection cross-section using Eqs.~\ref{22}-\ref{23}.
In this way the viable model parameter space for the dark matter
candidate $f_1$ is obtained.
Fig.~\ref{fig1}a is for the
case $m_2=$ 100 GeV while Fig.~\ref{fig1}b is for the case $m_2=$ 110 GeV.
The upper limit on $\sigma_{\rm SI}$ for different values of DM mass, obtained
from LUX DM direct search experiment, are also shown in Fig.~\ref{fig1}a-b by the
blue line for comparison. The red and green scattered regions as shown in
Fig.~\ref{fig1}a-b correspond to two choices of $v_s$=246 GeV and 500 GeV
respectively. From Fig.~\ref{fig1}a it can be observed that only the region
near the resonances of scalar bosons $h_1$ and $h_2$ is in agreement with the
upper limit on $\sigma_{\rm SI}$ predicted by LUX. It is also seen from
Fig.~\ref{fig1}a that the choice of $v_s$ do not alter the allowed range of
parameter space. Observation of Fig.~\ref{fig1}b yields that, apart from SM
Higgs resonance region ($m \sim m_1/2$) there exists another allowed range of
$m-\sigma_{\rm SI}$ parameter space in the vicinity of non-SM
scalar resonance ($m \sim m_2/2$). Note that variation of $m$ with $\sigma_{\rm
SI}$ depicted in Fig.~\ref{fig1}b depends only on the masses of scalar bosons
and does not suffer any significant change due to change in $v_s$.  The non-SM
Higgs signal strength $R_2$ (calculated using Eq.~\ref{9}) for the valid
$m-\sigma_{\rm SI}$ parameter space shown in
Figs.~\ref{fig1}a-b is very small and $R_2 < 0.2$.
\begin{figure}[h!]
\centering
\subfigure[]{
\includegraphics[height=7 cm, width=7 cm,angle=0]{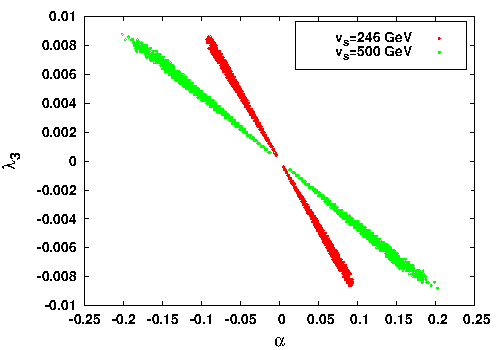}}
\subfigure []{
\includegraphics[height=7 cm, width=7 cm,angle=0]{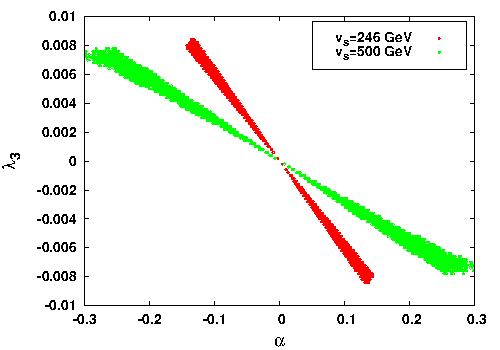}}
\caption{The valid model parameter space in $\lambda_3-\alpha$ (in deg) plane
obtained for the case of $m_2=100$ GeV (lef panel) and $m_2=$ 110 Gev (right
panel) satisfying the limits from vacuum stability, LHC findings, PLANCK DM
relic abundance and direct detection limits on $\sigma_{\rm SI}$ from LUX
experiment.}
\label{fig2}
\end{figure}
 
In this work, we assumed two values for VEV $v_s$ (246 GeV and 500 GeV) for the
hidden sector Higgs doublet $\Phi_{\rm HS}$. From Eq.~\ref{prob6}, we observe
that the mixing between the scalars $h_1$ and $h_2$ depends on the VEV of
$\Phi_{\rm HS}$ and $H$. Hence, the choice of $v_s$ may change the range of
available model parameter space. In Figs.~\ref{fig2}a-b, we plot the variation
of Higgs mixing angle $\alpha$ between $h_1$ and $h_2$ with $\lambda_3$ for
$m_2$=100 GeV and 110 GeV with $m_1$=125 GeV (mass of SM-like Higgs). Needless
to mention the region of $\alpha-\lambda_3$ space shown in Figs.~\ref{fig2}a-b
are consistent with the bounds form vacuum stability, SM Higgs signal strength 
from LHC, relic abundance of DM from PLANCK and limits on DM-nucleon scattering
cross-section from LUX direct DM search experiment. Plots in
Fig.~\ref{fig2} are produced using similar method we have applied previously to 
obtain viable model parameter space for Fig.~\ref{fig1}. However
$\alpha-\lambda_3$ plane in Fig. \ref{fig2} is further constrained by imposing
LUX DM direct detection bound. The plots in
Fig.~\ref{fig2}a are for the case when $m_1=$ 125 GeV and $m_2=$ 100 GeV
while plots in Fig.~\ref{fig2}b represent the allowed $\alpha-\lambda_3$
parameter space when $m_2=$ 110 GeV for the fixed value of $m_1=125$ GeV.
The green and blue regions in Fig.~\ref{fig2}a and Fig.~\ref{fig2}b correspond
to two different values for VEV of dark Higgs doublet, $v_s=$ 246 GeV and $v_s=$
500 GeV respectively. From Fig.~\ref{fig2}a ($m_2=$ 100 GeV case) one observes
that for both the considered values of VEV $v_s$, the mixing parameter
$\lambda_3$ remains small and is confined within the region $|\lambda_3|<0.01$.
For the case when $v_s=$246 GeV (the red region of Fig.~\ref{fig2}a), the limit
of mixing angle $\alpha$ ranges between $-0.1$ to $0.1$. However these range (of
mixing angle) varies within the limit $|\alpha|\leq0.2$ when $v_s$=500 GeV is
chosen (green region shown in Fig.~\ref{fig2}a). Study of the $\lambda_3-\alpha$
plots in Fig.~\ref{fig2}b (plotted for $m_2=$ 110 GeV) reveals that for both the
values of $v_s$ considered in Fig.~\ref{fig2}, the mixing parameter is small
($|\lambda_3|<0.01$). The mixing
angle $\alpha$ is bounded in the range $|\alpha|<0.15$ and $|\alpha|\leq0.30$
for $v_s=$ 246 GeV and 500 GeV respectively.
\subsection{Calculation of gamma ray signals from galactic centre and dwarf
galaxies}
\label{SS:indirect}
\begin{figure}[h!]
\centering
\includegraphics[height=7 cm, width=10 cm,angle=0]{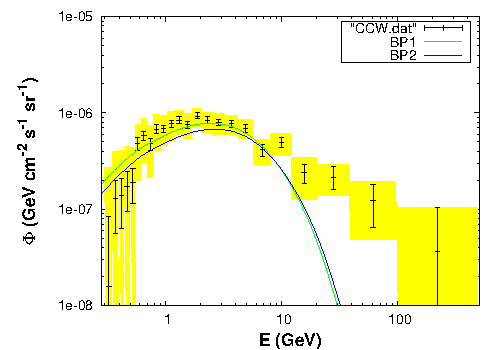}
\caption{Comparison of the GC $\gamma$-ray flux data from \cite{Calore:2014xka}
with those calculated for benchmark points in Table~\ref{tab1}.}
\label{fig3}
\end{figure}
\begin{table}
\begin{center}
\resizebox{\textwidth}{!}{
%\vskip 0.5 cm
\begin{tabular}{|c|c|c|c|c|c|c|c|}
\hline
    &          &           &        &            &            &              &  
      
                            \\ 
 BP &   $v_s$  &  $m_2$    &  $m$   & $\delta$   & $\gamma_2$   &$\sigma_{SI}$ &
$\langle
\sigma {\rm v}\rangle_{b\bar b}$ \\
    &  in GeV  & in GeV    & in GeV &            &            & in cm$^2$   &  
     
cm$^3$/s                      \\
\hline  
  BP1  &  246.0  & 100.242 &  50.0  & -4.86e-03  &  0.60e-06  &  2.89e-46   & 
1.98e-26  
                            \\ 
\hline 
 BP2   &  500.0  & 110.321 &  55.0  & -5.85e-03  &  0.43e-06  &  1.13e-46   & 
1.90e-26  
                            \\
\hline
\end{tabular}
}
\end{center}
\caption{Benchmark points obtained from the constrained model parameter space
in agreement with the bounds from vacuum stability, SM Higgs signal strength 
from LHC, DM relic density from PLANCK and LUX DM serach bounds on DM-nucleon
scattering cross-section.}
\label{tab1}
\end{table}
In this Section, we calculate the $\gamma$-ray flux from the galactic centre
and dwarf galaxies for the fermionic dark matter in framework of the present
model and compare our results with the experimental observations. For these
calculations we consider two benchmark points (BPs) from the restricted
parameter space that satisfy both theoretical and experimental bounds (mainly
vacuum stability, LHC constraints on SM Higgs signal, PLANCK results for relic
abundance and
direct detection limit on $m-\sigma_{\rm SI}$ from LUX) for two choices of
$h_2$ mass, mainly, $m_2$= 100 GeV and 200 GeV. In Table~\ref{tab1} we tabulate
the chosen BPs along with model  parameters. There are two chosen sets of
benchmark points in Table~\ref{tab1} and we denote them as BP1 and BP2. The GC
gamma ray flux is calculated using Eqs.~\ref{24}-\ref{26} for the BPs tabulated
in Table~\ref{tab1}. The annihilation cross-section $\langle
\sigma {\rm v}\rangle_{b\bar b}$ for the dark matter particle is calculated
using Breit-Wigner enhancment technique using Eqs.~\ref{30}-\ref{33} discussed
in Sec.~\ref{S:cons}. The gamma ray spectrum $\frac{dN}{dE}$ in Eq.~\ref{24} is
obtained from Ref. \cite{Cirelli:2010xx} for annihilation of DM into any
specific channel. The gamma ray spectra for BP1 and BP2 are then calculated for
the specified region of interest adopted from Ref.~\cite{Calore:2014xka}
($|l|\leq 20^0,~2^0\leq|b|\leq 20^0$) using NFW halo profile (with $\gamma=1.2$,
$\rho_{\odot}= 0.4~{\rm{GeV~cm^{-3}}}$). In Fig.~\ref{fig3}, we show the
calculated GC gamma ray flux (in GeV cm$^{-2}$ sr$^{-1}$) for our proposed DM
candidate with BP1 and BP2. We also show in Fig.~\ref{fig3} the CCW data for
comparison. Green and blue lines in Fig.~\ref{fig3} represent the calculated
$\gamma$-ray spectra for BP1 and BP2 respectively. Both the benchmarks points
are in agreement with the findings from GC gamma ray study presented in CCW
\cite{Calore:2014xka}. From Fig.~\ref{fig3} it can be observed that flux
calculated using the set BP1 ($m$=50 GeV) is in better agreement with the
findings from CCW analysis.

\begin{figure}[h!]
\centering
\includegraphics[height=7 cm, width=7 cm,angle=0]{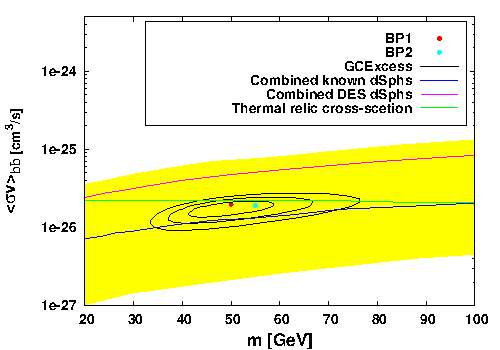}
\caption{The allowed range of $m-{\langle \sigma {\rm v}\rangle}$ space (for
annihilation into $b \bar b$) along with the bounds on ${\langle \sigma {\rm v}\rangle}$
(into $b \bar b$ channel only) obtained from GC $\gamma$ ray search results
CCW \cite{Calore:2014xka} and dwarf galaxies 
\cite{Ackermann:2015zua,Drlica-Wagner:2015xua} compared with
the same obtained from benchmark points in Table~\ref{tab1}.}
\label{fig4}
\end{figure}

We now further 
investigate how well the DM candidate in our model can explain the observed 
extragalactic $\gamma$-ray signatures from various dwarf galaxies. 
From their six years observations on 15 dwarf galaxies, the Fermi-LAT experiment
did not obtain any significant excess of $\gamma$-rays. Fermi-LAT collaboration
\cite{Ackermann:2015zua} however in a recent work provides combined bound on DM
mass and thermally averaged DM annihilation cross-section into SM particles for
these 15 dSphs. A similar bound in $m-{\langle \sigma {\rm v}\rangle}_f$ plane is
also presented recently in an another work \cite{Drlica-Wagner:2015xua} for
eight new dSphs jointly by Fermi-LAT and DES collaboration. In this work we
calculate thermally averaged annihilation cross-section of DM annihilating into
SM sector in our model and compare them with experimental results given by
\cite{Ackermann:2015zua,Drlica-Wagner:2015xua}. In Fig.~\ref{fig4}, we plot the
bounds on DM annihilation cross-section ${\langle \sigma {\rm v}\rangle}_{b \bar b}$
(for the annihilation channel DMDM$\rightarrow b \bar b$) with dark matter mass
$m$ obtained from galactic \cite{Calore:2014xka} and extragalactic 
\cite{Ackermann:2015zua,Drlica-Wagner:2015xua} $\gamma$-ray search experiments.
We calculate the variations of the same plotted in Fig.~\ref{fig4} for the 
benchmark points BP1 (for $m_2=$ 100 GeV) and BP2 (for $m_2=$ 110 GeV) 
considered in our model.

Black contours shown in Fig.~\ref{fig4} are the 1$\sigma$, 2$\sigma$ and
3$\sigma$ contours given by the CCW \cite{Calore:2014xka} analysis of GC gamma
ray excess observations. The blue line in Fig.~\ref{fig4} describes the bounds
in $m-{\langle \sigma {\rm v}\rangle}_{b \bar b}$ plane given by the analysis of 
gamma rays from previously discovered 15 dSphs and they are adopted from 
\cite{Ackermann:2015zua}. Also shown in Fig.~\ref{fig4}, the yellow band which is 
the 95\% confidence limit (C.L.) region adopted from the analysis in 
Ref.~\cite{Ackermann:2015zua} for DM annihilation into $b\bar b$. The combined
bounds on ${\langle \sigma {\rm v}\rangle}_{b \bar b}$ for different DM mass
$m$ from a recent study of the newly discovered 8 DES dwarf galaxies
\cite{Drlica-Wagner:2015xua} are given by the pink
coloured line in Fig.~\ref{fig4}. The green horizontal line in Fig.~\ref{fig4} shows the 
annihilation cross-section for thermal dark matter that may yield  the right
DM relic abundance obtained from the PLANCK experiment.

From Fig.~\ref{fig4} one readily observes that the calculated values
of ${\langle \sigma {\rm v}\rangle}_{b \bar b}$ for the benchmark points BP1 
and BP2 in our model broadly agrees with the 1$\sigma$, 2$\sigma$ and 
3$\sigma$ allowed regions in $m-{\langle \sigma {\rm v}\rangle}_{b \bar b}$
plane obtained from the experimental results.
This can also be noted from Fig.~\ref{fig4} that these benchmark points are
consistent the combined limit from DES dwarf satellite
data and falls within the 95\% C.L. limit predicted by Fermi-LAT for 15 dSphs.
Also the calculated values of ${\langle \sigma {\rm v}\rangle}_{b \bar b}$ for
the benchmark points considered in our work lie below the upper bound on thermal DM annihilation
cross-section. Hence, DM fermion in the present model can account for the galactic centre excess in
$\gamma$-ray and is also consistent with the bounds on gamma ray flux from
Milky-Way dwarf satellite galaxies. 
\begin{figure}[h!]
\centering
\subfigure[]{
\includegraphics[height=4.5 cm, width=4.5 cm,angle=0]{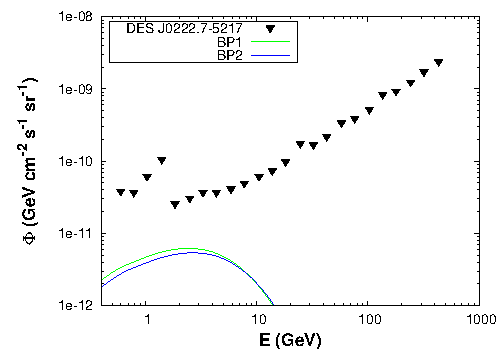}}
\subfigure []{
\includegraphics[height=4.5 cm, width=4.5 cm,angle=0]{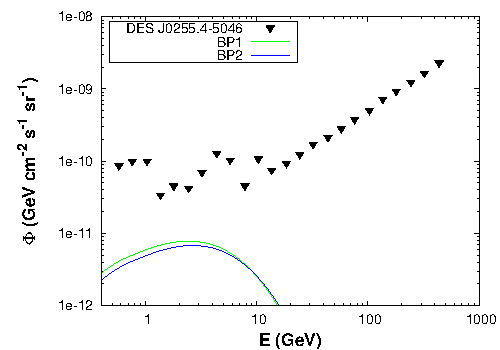}}
\subfigure[]{
\includegraphics[height=4.5 cm, width=4.5 cm,angle=0]{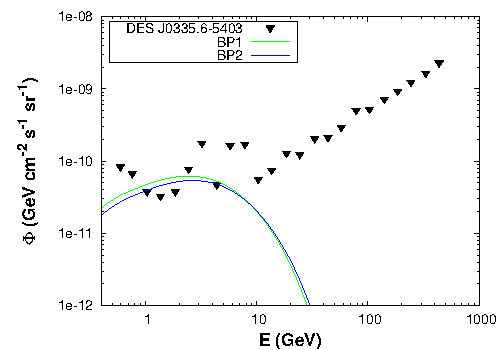}}
\subfigure []{
\includegraphics[height=4.5 cm, width=4.5 cm,angle=0]{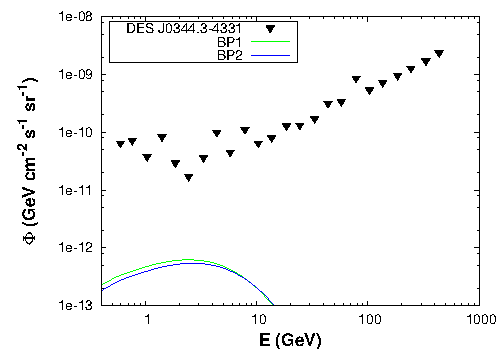}}
\subfigure[]{
\includegraphics[height=4.5 cm, width=4.5 cm,angle=0]{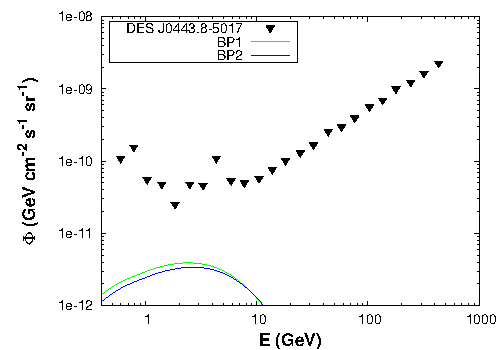}}
\subfigure []{
\includegraphics[height=4.5 cm, width=4.5 cm,angle=0]{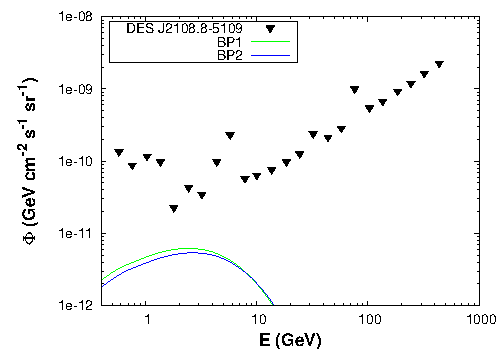}}
\subfigure[]{
\includegraphics[height=4.5 cm, width=4.5 cm,angle=0]{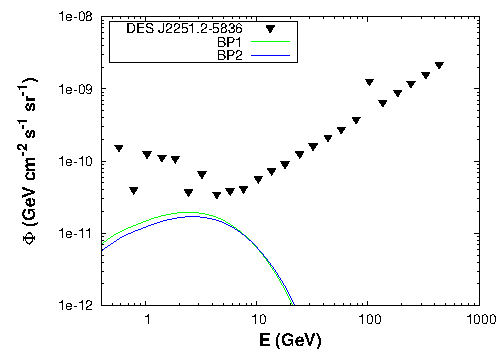}}
\subfigure []{
\includegraphics[height=4.5 cm, width=4.5 cm,angle=0]{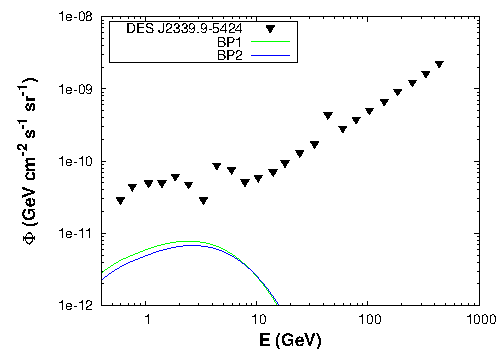}}
\caption{Comparison of the observed upper bound on $\gamma$-ray flux for 8
DES dSphs with the calculated $\gamma$-ray flux from BP1 and BP2 tabulated
in Table~\ref{tab1}.}
\label{fig5}
\end{figure}

We now calculate the gamma ray flux for 8 new dwarf satellite galaxies
discovered by the DES experiment for the hidden sector fermionic dark matter
candidate proposed in this work. These calculations are performed with
each of the benchmark parameter sets BP1 and BP2 given in Table~\ref{tab1}.
The Gamma ray flux for each of these 8 dSphs in the work
\cite{Drlica-Wagner:2015xua} is computed using Eq.~\ref{27} and the values of
the $J$ factors (Eq.~\ref{28}) for each of the eight dSphs adopted from
Ref.~\cite{Drlica-Wagner:2015xua}. In Ref.~\cite{Drlica-Wagner:2015xua} these
$J$ factors are estimated by integrating the dark matter density (adopting NFW
halo profile for DM density distribution) along the line of sight over a solid
angle $\Delta\Omega=2.4\times 10^{-4}$ sr$^{-1}$. As previously mentioned the
gamma ray spectrum $\frac{dN}{dE}$ is also obtained from
Ref.~\cite{Cirelli:2010xx} for this calculation. The calculated flux for each of
the eight dSphs are shown in eight plots (a-h) of Fig.~\ref{fig5}. Also shown in
each of the eight plots of Fig.~\ref{fig5}, the respective upper bounds of the
flux given by the experimental observations of gamma rays from each of the eight
dSphs. These are shown as red coloured points while the computed flux in this
work for the respective dSphs are given by continuous lines in Fig.~\ref{fig5}.
The green and blue continuous lines in each of the plots (a-h) of
Fig.~\ref{fig5} correspond to the calculated flux using the benchmark points
BP1 and BP2 respectively. It is clear from Fig.~\ref{fig5} that the fluxes
calculated, assuming the annihilation of the DM candidate in our proposed
model, for all the eight dSphs do not exceed the upper limit of $\gamma$ flux
set by the experimental observations of DES collaboration.
 
\begin{figure}[h!]
\centering
\includegraphics[height=7 cm, width=7 cm,angle=0]{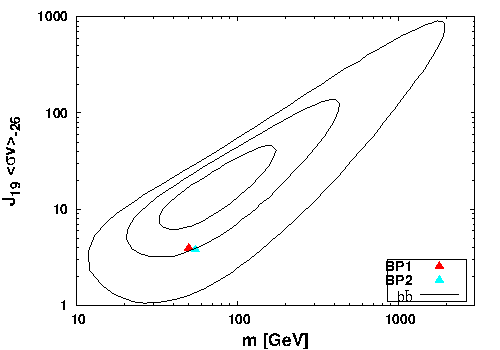}
\caption{Benchmark points BP1 and BP2 compared with the allowed region of 
model parameter space shown in  $m-J_{19}\langle \sigma {\rm v} \rangle_{-26}$
plane
obtained from \cite{Geringer-Sameth:2015lua}}
\label{fig6}
\end{figure}

Besides the 15 dwarf galaxies investigated earlier and the eight other recently
explored dwarf galaxies, one more dwarf galaxy namely Reticulum 2 (Ret2)
has been probed very recently. Geringer-Sameth et. al. \cite{Geringer-Sameth:2015lua},
after an analysis of observed gamma rays from Ret2 dwarf galaxy reported an
excess of gamma ray emission from Ret2. From their analysis of Ret2 data
%\cite{Atwood:2009ez} 
Geringer-Sameth et. al. provide different C.L. allowed contours in 
$m-J_{19}\langle \sigma {\rm v}\rangle_{-26}$ plane where $m$ is the mass of the dark
matter and $J_{19}\langle \sigma {\rm v}\rangle_{-26}$ is the product of the $J$
factor in the units of $10^{19}$ GeV$^2$ cm$^{-5}$ and thermal
averaged product $\langle
\sigma {\rm v}\rangle$ of annihilation cross-section and relative velocity in the
units of $10^{-26} {\rm cm^3 s^{-1}}$ for various final state SM channels. As
mentioned earlier in this work DM candidate primarily annihilates into $b \bar
b$, only the contours for the DM pair annihilation into $b \bar b$ channel
are adopted. For the present dark matter model with the constrained
parameter space discussed earlier we compute the quantity $J_{19}\langle \sigma
{\rm v}\rangle_{-26}$ for different dark matter mass $m$ annihilating into $b \bar b$
channel. However the value of the $J$ factor for Ret2 has been adopted from
\cite{Geringer-Sameth:2015lua}. In their work Geringer-Sameth et. al.
\cite{Geringer-Sameth:2015lua} estimated the $J$ values by performing line of
sight integral over a circular region with angular radius $0.5^0$ surrounding
the dwarf and over a solid angle $\Delta\Omega=2.4\times 10^{-4}$ sr$^{-1}$. All
these calculations are performed for two values of non-SM scalar mass accounted
in the present model namely $m_2=$ 100 GeV and $m_2=$ 110 GeV. The results are
presented for the two benchmark points BP1 and BP2 corresponding to the 
calculations with $m_2=$ 100 GeV and $m_2=$ 110 GeV are shown in red and skyblue
points in Fig.~\ref{fig6}. In Fig.~\ref{fig6}, the contours from the 
experimental data analysis by Geringer-Sameth et. al. are given for comparison. In 
Fig.~\ref{fig6} the contours for 68\%, 95\% and 99.7\% C.L.  are shown in
black coloured lines in increasing order of area enclosed by each contour. 
The valid regions of  $m-J_{19}\langle \sigma {\rm v}
\rangle_{-26}$ plane in our model (calculated for DM annihilating into $b \bar
b$ pair) are presented by green coloured patches in both the plots of
Fig.~\ref{fig6}. From Fig.~\ref{fig6}, it can be easily observed that
$J_{19}\langle \sigma {\rm v} \rangle_{-26}$ in the present model calculated for DM
annihilating into $b \bar b$ channel (for benchmark points with $m_2$=100 GeV and
110 GeV) is within the $3\sigma$ C.L. limit. Hence fermionic DM candidate in the
present framework can also explain the observed excess in $\gamma$-ray from
Ret2.

\section{Discussions and Conclusions}
\label{S:con}

In this work, we have proposed the existence of a hidden sector which obeys
a local SU(2)$_{\rm H}$ and a global U(1)$_{\rm H}$ gauge symmetries.
In order to introduce fermions which are charged under this
${\rm SU}(2)_{\rm H}$ gauge group one should have at least
two fermion doublets in order to avoid ``Witten anomaly".
The particle and the antiparticle of these dark fermions
are different as they possess equal and opposite U(1)$_{\rm H}$
charges. Similar to the usual Higgs doublet in
the visible sector, this hidden sector also has an
SU$(2)_{\rm H}$ scalar doublet $\Phi$ which however does not have any
U(1)$_{\rm H}$ charge. The ${\rm SU}(2)_{\rm H}$ gauge
symmetry breaks spontaneously when the neutral component of
the scalar doublet $\Phi$ gets a VEV and thereby
generates masses of all the dark gauge bosons ($A^{\prime}_{\mu}$)
and dark fermions ($f_i$).
Since the dark sector fermions interact among themselves through the dark
gauge bosons, therefore all the heavier fermions as well as the dark
gauge bosons can decay into the lightest fermion and hence the lightest
fermion in this dark sector can be treated as a particle for
the viable dark matter candidate. In fact in this model this
lightest fermion is the only dark matter candidate.
The dark fermions and dark gauge bosons do not mix with
the SM fermions and gauge bosons due to the non abelian
nature of two SU(2) groups. However,
the dark sector scalar field can interact with the
SM Higgs like scalar in the visible sector and only through this
interaction two sectors are mutually connected. 

We therefore test the viability of the present model
by using theoretical and experimental constraints on
the relevant model parameters, such as vacuum stability
conditions, bounds on relic abundance of DM from PLANCK
experiment, direct detection limits on DM-nucleon
scattering cross-section from LUX experiment. LHC bounds
on signal strength and invisible decay width of the SM Higgs,
are also used to constrain the parameter space.
From such analyses we find that only a small region of the
parameter space near the scalar resonances (when
$m\simeq \frac{m_1}{2}$ and $\frac{m_2}{2}$), is consistent
with the current experimental bounds. Study of the model parameters,
thus constrained, shows that the mixing between the two scalars ($h_1$, $h_2$)
of the model is very small (mixing angle $\alpha \le 0.3$ deg) and depends on
the VEV of the dark scalar doublet ($\Phi$). With the allowed regions
of parameter space, thus obtained, for the present DM candidate
(dark fermion $f_1$) we compute the gamma ray flux
from the GC region. While calculating the gamma ray flux
from GC we have used Breit-Wigner enhancement mechanism
for the computation of DM annihilation cross-section
into the $b \bar b$ final state ($f_1 \bar{f_1} \rightarrow b \bar b$).
These computational results are then compared with the experimental
analyses of the Fermi-LAT GC gamma ray flux data
considering the dark matter at the GC primarily pair
annihilates into $b \bar b$ channel. Our proposed DM candidate
can indeed explain the results from these experimental analyses.

In search of indirect evidence of dark matter from astrophysical sources,
the gamma rays from various dwarf satellite galaxies are also explored for
possible signature of excess gamma rays from these sites. To this end 15 such
dwarf galaxies have earlier been investigated and more recently the gamma ray
observation is also reported from eight more newly discovered dSphs.
From the analyses of these observational results different C.L. bounds have been given
in the parameter space of ${\langle \sigma {\rm v}\rangle}_{b \bar b}-m$ plane. We
compare our computational results with these experimental bounds and found that
the $\gamma$-rays that the DM candidate in our model produce on pair annihilation
can simultaneously satisfy the observational results from GC and dwarf galaxies.
We also demonstrate that the calculated fluxes in our model for each of the 
recently discovered eight dwarf galaxies lie below the corresponding upper
limits of the fluxes obtained from the observational results of these dwarf
galaxies. We further demonstrate that our calculations are also in good agreement
with the analysis of Ret2 dwarf galaxy observations.  

Our work clearly demonstrate that the dark matter candidate proposed in this
work is a viable one to explain the $\gamma$-rays from both the GC region and
dwarf galaxies simultaneously. However the dark matter can also pair annihilate
into fermion-antifermion pairs and there are experiments such as 
AMS-02 that look for the excess of $e^+/e^-$ or $p \bar p$ in cosmos. 
In a recent work, AMS-02 collaboration have reported their first measurement of
$p/\bar p$ flux \cite{ams02}. A model independent analysis of this AMS-02
$p/\bar p$ data is performed by Jin et.al. \cite{Jin:2015sqa}. In this work
\cite{Jin:2015sqa}, the upper limits in $\langle \sigma {\rm v}\rangle$ value for DM
annihilation into SM particles (quarks and gauge bosons) for different
considered DM halo profiles (NFW, Isothermal, Moore) are obtained. The
analysis presented in the work \cite{Jin:2015sqa} also considered four different
propagation models namely conventional, MED, MIN and MAX\footnote {For further
studies see \cite{Jin:2015sqa} and references therein.}. We have also checked
that the DM in our model satisfies upper bound on $\langle \sigma {\rm v} \rangle_{b
\bar b}$ given in Ref.~\cite{Jin:2015sqa} when NFW profile is considered. This
is found to be true for both the cases of dark sector scalar mass $m_2=$ 100
GeV and $m_2=$ 110 GeV.  Hence, fermionic dark matter explored in the present
model can serve as a potential candidate for dark matter. Upcoming results from
LHC as also DM direct and indirect search experiments may provide stringent
limits on the available model parameter space.

{\bf Acknowledgments} : A.D. Banik and A. Biswas would like to thank P.B. Pal
for useful discussions. A.D.B. and A.B. also acknowledge the Department
of Atomic Energy, Govt. of India for financial support.
        
\vskip 5mm
\noindent {\bf Appendix A}
\vskip 2mm 
Coupling between the scalars $h_1$ and $h_2$ are given as follows
\bea
&&\lambda_{111}=\lambda_1 v c_{\alpha}^3 - \lambda_2 v_s s_{\alpha}^3
+\frac{1}{2}\lambda_3(v c_{\alpha}s_{\alpha}^2 - v_s s_{\alpha}c_{\alpha}^2)\, , \nonumber \\
&&\lambda_{222}=\lambda_1 v s_{\alpha}^3 + \lambda_2 v_s c_{\alpha}^3
+\frac{1}{2}\lambda_3(v s_{\alpha}c_{\alpha}^2 + v_s c_{\alpha}s_{\alpha}^2)\, , \nonumber \\
&&\lambda_{211}=3(\lambda_1vc_{\alpha}^2s_{\alpha}-\lambda_2v_ss_{\alpha}^2c_{\alpha})
+\frac{1}{2}\lambda_3(v_s(c_{\alpha}^3-2s_{\alpha}^2c_{\alpha})+
v(s_{\alpha}^3-2c_{\alpha}^2s_{\alpha}))\, , \nonumber \\
&&\lambda_{122}=3(\lambda_1vs_{\alpha}^2c_{\alpha}-\lambda_2v_sc_{\alpha}^2s_{\alpha})
+\frac{1}{2}\lambda_3(v_s(-s_{\alpha}^3+2c_{\alpha}^2s_{\alpha})+
v(c_{\alpha}^3-2s_{\alpha}^2c_{\alpha}))\, . \nonumber
\eea

\end{document}